\documentclass[twocolumn,superscriptaddress,aps,floatfix,preprintnumbers,amsmath,amssymb,prl,footinbib,10pt]{revtex4-1}
\usepackage[titletoc,toc,title]{appendix}
\usepackage[english]{babel}
\usepackage[hidelinks]{hyperref}
\usepackage[normalem]{ulem}
\usepackage{adjustbox}
\usepackage{amsfonts,amsmath,amssymb}
\usepackage{bm}
\usepackage{enumitem}
\usepackage{epsfig}
\usepackage{cancel}
\usepackage{centernot}
\usepackage{color}
\usepackage{comment}
\usepackage{contour}
\usepackage{flushend}
\usepackage{footmisc}
\usepackage{graphics}
\usepackage{graphicx}
\usepackage{mathrsfs}
\usepackage{mdframed}
\usepackage[normalem]{ulem}
\usepackage{pifont}
\usepackage{relsize}
\usepackage{shadowtext}
\usepackage{slashed}
\usepackage{soul}
\usepackage{subfigure}
\usepackage{tcolorbox}
\usepackage{textgreek}
\usepackage{titlesec}
\usepackage{titletoc}
\usepackage{verbatim}
\usepackage{xcolor}
\usepackage{xfrac}
\contourlength{0.2em}
\usepackage{orcidlink}
\usepackage{ragged2e} 

\definecolor{zima_blue}{HTML}{1393C1}
\hypersetup{setpagesize=false,bookmarksnumbered=true,bookmarksopen=true,colorlinks=true,linkcolor=zima_blue,urlcolor=zima_blue,citecolor=zima_blue,linktocpage=false}

\begin{document}

\title{Big Bang Nucleosynthesis confronts Domain Walls and First-Order Phase Transitions as the Pulsar Timing Array Explanations}
\author{Fatemeh Elahi}
\email{felahi@uni-mainz.de}
\affiliation{PRISMA$^{++}$ Cluster of Excellence $\&$ Mainz Institute for Theoretical Physics, Johannes Gutenberg University, 55099 Mainz, Germany}
\author{Diego Fortini}
\email{dfortini@students.uni-mainz.de}
\affiliation{PRISMA$^{++}$ Cluster of Excellence $\&$ Mainz Institute for Theoretical Physics, Johannes Gutenberg University, 55099 Mainz, Germany}
\author{Yann Gouttenoire}
\email{yann.gouttenoire@iap.fr}
\affiliation{Institut d’Astrophysique de Paris (IAP), CNRS, Sorbonne Universit\'e, FR-75014, France}
\affiliation{PRISMA$^{++}$ Cluster of Excellence $\&$ Mainz Institute for Theoretical Physics, Johannes Gutenberg University, 55099 Mainz, Germany}
\author{Nicholas Leister}
\email{nleister@uni-mainz.de}
\affiliation{PRISMA$^{++}$ Cluster of Excellence $\&$ Mainz Institute for Theoretical Physics, Johannes Gutenberg University, 55099 Mainz, Germany}
\author{Pedro Schwaller}
\email{pedro.schwaller@uni-mainz.de}
\affiliation{PRISMA$^{++}$ Cluster of Excellence $\&$ Mainz Institute for Theoretical Physics, Johannes Gutenberg University, 55099 Mainz, Germany}

\begin{abstract}
Pulsar timing arrays (PTAs) have detected a nanohertz gravitational-wave background, often attributed to annihilating domain walls (DWs) or first-order phase transitions (FOPTs). Their reheating imprints fluctuations in the baryon-to-photon ratio. These so-called baryon isocurvature fluctuations $S_B$ survive until Big Bang Nucleosynthesis, whose yields depend nonlinearly on it. Modest $S_B$ fluctuations can therefore alleviate the current deuterium tension in parts of the PTA preferred regions and therefore strengthen their interpretation. Overabundant DWs and too strong FOPTs overproduce $D/H$ and are ruled out.
\end{abstract}
\preprint{MITP-26-047}
\maketitle
 
\textit{\textbf{Introduction.}}
Pulsar timing arrays (PTAs) have reported evidence for a stochastic gravitational-wave (GW) background at nanohertz frequencies~\cite{NANOGrav:2023gor,EPTA:2023fyk,Reardon:2023gzh,Xu:2023wog}. Its amplitude and comparatively steep spectrum are difficult to reconcile with the expected astrophysical population of inspiraling supermassive black-hole binaries~\cite{NANOGrav:2023hfp,EPTA:2023xxk}, and have instead motivated a broad range of cosmological interpretations~\cite{NANOGrav:2023hvm,Ellis:2023oxs,Madge:2023dxc,Figueroa:2023zhu}. Among the most studied are energetic processes in the early Universe, in particular the annihilation of domain walls (DWs)~\cite{Zeldovich:1974uw,Vilenkin:1984ib,Saikawa:2017hiv,Ferreira:2022zzo,Gouttenoire:2023ftk} and cosmological first-order phase transitions (FOPTs)~\cite{Caprini:2015zlo,Caprini:2019egz,Xue:2021gyq,Ratzinger:2020koh,Xue:2021gyq,NANOGrav:2021flc,Gouttenoire:2023bqy}, whose GWs peak in the PTA band if they occur around the epoch of quark-gluon confinement.

Reproducing the observed amplitude is nonetheless demanding. In both scenarios the emission is controlled by large, spatially inhomogeneous fluctuations in the energy density of the source: DW networks release their energy in Hubble-sized patches when they annihilate, whereas FOPTs complete through the stochastic nucleation of bubbles. The strong, slow transitions that maximize the PTA signal are therefore precisely those with the largest source inhomogeneities.

These same fluctuations are imprinted on the primordial plasma. Local energy injections that do not change the baryon number but locally produce photons generate baryon-to-photon isocurvature $S_B$, i.e. spatial fluctuations in the respective ratio $\eta_B\equiv n_B/n_\gamma$. Because light-element yields depend nonlinearly on the local $\eta_B$, even fluctuations with vanishing spatial mean do not average out but shift the mean abundances, raising in particular the deuterium yield~{\cite{Applegate:1985qt,Applegate:1987hm,Alcock:1987tx,Jedamzik:1994ux,Kurki-Suonio:1996wfr,Barrow:2018yyg,Inomata:2018htm,Bagherian:2025puf}}. Intriguingly, standard BBN currently underpredicts the observed $D/H$ by {up to} $1.7\sigma${, depending on how the deuterium-burning rates are fitted}~\cite{Cooke:2017cwo,Pitrou:2020etk,Pitrou:2021vqr,Launders:2026ciu,Pisanti:2020efz,Yeh:2020mgl,Schoneberg:2024ifp,Chen:2025oyj}, a possible hint of new physics~{\cite{Deal:2021kjs,Burns:2022hkq,Froustey:2024mgf}}. It sharpens further if the Hubble tension has an early-time solution, since such solutions generically prefer a higher baryon density $\Omega_B$~\cite{Giovanetti:2026aku}, and a non-standard expansion history during deuterium burning has been proposed to resolve it~\cite{Poulin:2026ltf,McKeen:2024voa,Cook:2025gra}. A baryon isocurvature naturally raises the predicted abundance toward the data, and therefore supports the DW and FOPT interpretations of the PTA signal in parts of the parameter space. {Whether it does so, or merely constrains them, depends on the nuclear rates: the \texttt{PRIMAT} prediction~\cite{Pitrou:2018cgg} lies $1.7\sigma$ below the measured $D/H$~\cite{Cooke:2017cwo}, whereas \texttt{PArthENoPE}~\cite{Gariazzo:2021iiu} already agrees and leaves only an upper bound, so we carry both throughout.}

In this \textit{Letter} we describe these fluctuations as a baryon isocurvature perturbation $S_B$ within cosmological perturbation theory, conserved on all PTA relevant scales and surviving until BBN. The implications are twofold: 1) A modest $S_B$ therefore relieves the $1.7\sigma$ deuterium tension across part of the preferred region for PTA. 2) Too strong FOPTs and overabundant DW overproduce $S_B$ and raise $D/H$ well above the observed value. These regions are thereby excluded, which currently yields the tightest bounds on the DW and FOPT interpretation of the PTA signal. {Using precision BBN to constrain sources that leave $\eta_B$ spatially varying goes back to the quark-hadron transition~\cite{Applegate:1985qt,Alcock:1987tx} and to electroweak baryogenesis~\cite{Fuller:1993sp,Heckler:1994uu,Brandenberger:1994fe}, and has recently been extended to a range of baryogenesis models~\cite{Azatov:2026sdm}. Those sources vary the baryon number, whereas here the photon density varies at fixed baryon number, with the mixed case treated in App.~\ref{app:modeldep}. Ref.~\cite{Bagherian:2025puf} established this bound and first flagged the PTA interpretations as targets, leaving their amplitude to a dedicated calculation. We provide it.} Detailed derivations are deferred to the Supplemental Material.
 
\textit{\textbf{From source fluctuations to isocurvature.}}
We quantify the  inhomogeneity in the baryon-to-photon ratio
$\eta_B(\mathbf{x})=n_B(\mathbf{x})/n_\gamma(\mathbf{x})$ by the baryon isocurvature perturbation $S_B(\mathbf{x})$, defined through~\footnote{Technically, we use the
net baryon-to-photon isocurvature with $\delta_B\equiv\delta_{B-\bar{B}}$. In the Supplemental Material we show that $S_B\simeq S_{B+\bar{B}}\simeq S_{B-\bar{B}}$ after baryon-anti-baryon freeze-out.}
\begin{equation}
S_B\equiv\frac{\delta\eta_B}{\bar\eta_B}
=\frac{\delta n_B}{n_B}-\frac{\delta n_\gamma}{n_\gamma}
=\delta_B-\frac34\delta_\gamma ,
\end{equation}
where $\delta_i\equiv\delta\rho_i/\bar\rho_i$ denotes the density contrast. The factor $3/4$ comes from the difference in temperature scaling $n_\gamma\propto\rho_\gamma^{3/4}$and $n_B\propto \rho_B$.
Assuming the baryon asymmetry was established prior to energy injection and remains unperturbed ($\delta_{\mathrm{B}} \simeq 0$)~\footnote{If the DW annihilation or FOPT itself drives baryogenesis, spatial variations in the energy injection will instead induce correlated fluctuations in the baryon number density $\delta_{\mathrm{B}}$, see e.g.~\cite{Heckler:1994uu,Megevand:2004ry,Daido:2015gqa,Bagherian:2025puf,Azatov:2026sdm}, modifying the initial amplitude and phase of the isocurvature mode. We discuss this in more detail in App.~\ref{app:modeldep}}\nocite{Megevand:2004ry,Daido:2015gqa}, overheated regions develop a local photon overdensity $\delta_\gamma$. Parameterizing the normalized energy injection of the GW source as
\begin{equation}
\label{eq:alpha}
\alpha_S \equiv \frac{\rho_S}{\rho_{\mathrm{SM}}},
\end{equation}
and  the  source density contrast as $\delta_S$, the inhomogeneous heating yields a photon density fluctuation $\delta_\gamma \simeq \frac{\alpha_S}{1 + \alpha_S}\delta_S$.
With $S_B=\delta_B-\tfrac34\delta_\gamma$ this yields
\begin{equation}
\label{eq:Sb_estimate}
S_B\simeq-\frac34\,\frac{\alpha_S}{1+\alpha_S}\,\delta_{\rm S}, \qquad S=\{\rm DW, FOPT\}.
\end{equation}
The same large fluctuations $\delta_S$ and large energy fraction $\alpha$ that make DWs and FOPTs efficient GW sources therefore unavoidably generate a sizable baryon isocurvature.
 
\textit{\textbf{Isocurvature evolution.}}
The evolution of the baryon isocurvature is well captured by cosmological
perturbation theory, which provides a clear framework to follow the fluctuations in the baryon-to-photon ratio from their production to BBN. We work at linear order and choose the Newtonian gauge
\begin{equation}
ds^2=a^2[-(1+2\Psi)d\eta^2+(1-2\Phi)\delta_{ij}dx^idx^j].
\end{equation}
Different particle species $a$ enter through perturbations of the total energy-momentum tensor $T^{\mu}{}_{\nu}\equiv\sum_a T^{\mu}{}_{\nu}^{(a)}$~\cite{Baumann:2022mni},
\begin{align}
T^{\mu}{}_{\nu} =
\begin{pmatrix}
-(\bar\rho+\delta\rho) & (\bar\rho+\bar P)\,v_j
\\[6pt]
-(\bar\rho+\bar P)\,v^i & (\bar P+\delta P)\,\delta^{i}{}_{j}+\Pi^{i}{}_{j}
\end{pmatrix},
\end{align}
which define the total density contrast $\delta$, velocity $v$, pressure perturbation $\delta P$ and anisotropic stress $\Pi$. The total $T^\mu{}_\nu$ sources $\Phi,\Psi$ through the Einstein equations. At linear order and in Fourier space with wavevector $k$, the Einstein equations and covariant energy-momentum conservation $\nabla_{\mu} T^{\mu\nu}=0$ yield the continuity and Euler equations of the
different fluids
\begin{align}
    \frac{{\rm d} \delta_a}{\rm d \eta}=\;&(1+\omega_a)\left(kv_a+3\frac{\rm d \Phi}{\rm d \eta}\right),\\
\frac{{\rm d} v_a}{\rm d \eta}=\;&-\mathcal{H}(1-3\omega_a)v_a-\frac{\omega_a}{1+\omega_a}k\delta_a+\frac{2}{3}k\Pi_a-k\Psi \notag 
\\&-\sum_b\Gamma_{ab}(v_a-v_b).
\end{align}
Here $\Gamma_{ab}$ denotes the interaction rate between species and
$\Pi_a\simeq v_ak/\left(\sum_b\Gamma_{ab}\right)$ the anisotropic stress from
free-streaming. The system is closed by the shear equation from the linearized Einstein equations
\begin{equation}
    \Phi-\Psi = \frac{3\mathcal H^2 \sum_a\Omega_a(1+\omega_a)\Pi_a}{k^2},
\end{equation}
where $\Omega_a\equiv\bar\rho_a/\bar\rho$. The photon and baryon \footnote{This already assumes that quarks and gluons have hadronized. Before the QCD phase transition our conclusions remain the same under the replacement $B\rightarrow q$.} continuity equations,
$\delta_\gamma'=\tfrac43 kv_\gamma+4\Phi'$ and $\delta_B'=kv_B+3\Phi'$, share the same curvature term, which cancels in $S_B=\delta_B-\tfrac34\delta_\gamma$:
\begin{equation}
\label{eq:Sb_evolution}
S_B'=k\,(v_B-v_\gamma).
\end{equation}
The baryon-to-photon ratio is therefore conserved whenever the fluids are tightly coupled. Before recombination, electrons, baryons and photons are locked together by Coulomb and Compton scattering, which gives (see App.~\ref{app:perturbation} for derivation)
\begin{equation}
  \frac{S_B'}{k} 
  \simeq
  \frac{k}{\Gamma_{\gamma e}+\Gamma_{\gamma B}}\,  \mathcal{O}\big(\delta_\gamma,v_B,\dots\big)\simeq\frac{k}{10^{21}~\rm Mpc^{-1}}
\end{equation}
as we confirm with a full numerical evolution (see App.~\ref{app:perturbation} and Ref.~\cite{Applegate:1988gb,Alcock:1990bb,Jedamzik:1993tcf,Jedamzik:1993dc}). Any baryon isocurvature associated with PTA scale inhomogeneities therefore survives intact until BBN. Neutrino free-streaming leaves these conclusions unchanged, cf. App.~\ref{app:perturbation} (see also Refs.~\cite{Heckler:1993nc,Jedamzik:1993tcf}). As the neutrinos begin to free stream they initially drag the photons along and erase the photon overdensities. The baryon--photon fluid, however, remains tightly coupled ($v_B\simeq v_\gamma$), so any outflow of photons is matched by an outflow of baryons. Erasing a photon overdensity thus creates a compensating underdensity in the (net) baryon number, and the baryon isocurvature is conserved across the neutrino free-streaming transition~\footnote{The same holds when the higher moments of the neutrino Boltzmann hierarchy become relevant around neutrino decoupling. The tight baryon--photon coupling still enforces an exact compensation between the baryon and photon energy densities, so even as the free-streaming neutrinos and their higher moments erase the fluctuations in the individual fluids, the baryon isocurvature $S_B$ is preserved. After neutrino decoupling, the neutrino should be evolved separately.}.

\textit{\textbf{BBN constraint.}}
Because BBN depends nonlinearly on $\eta_B$, the light-element abundances are sensitive not only to the mean baryon-to-photon ratio $\bar\eta_B$, but also to its spatial fluctuations~{\cite{Jedamzik:1994ux,Barrow:2018yyg,Inomata:2018htm,Bagherian:2025puf}}. For a zero-mean isocurvature perturbation, the linear contribution averages to zero, while the variance $\langle S_B^2\rangle$ does not. Deuterium is particularly sensitive because its abundance decreases steeply with $\eta_B$.
Since $D$ and $H$ homogenize after BBN and are measured independently the relevant observable is the ratio of spatially averaged number densities, $\langle n_D\rangle/\langle n_H\rangle$, rather than the volume average of the local ratio $n_D/n_H$~{\cite{Bagherian:2025puf}}. {Because the response is quadratic and positive, isocurvature can only raise $\langle D\rangle/\langle H\rangle$, never lower it. A prediction that falls below the measurement is therefore raised toward it, whereas one that already agrees can only be pushed above it. The predicted deuterium abundance also depends on the treatment of the nuclear reaction rates, in particular on which data are included and how the cross sections are fitted~{\cite{Schoneberg:2024ifp}}. Recent analysis~\cite{Launders:2026ciu}, that used a different fitting procedure, supports the central values predicted by \texttt{PRIMAT}~\cite{Pitrou:2018cgg}. Since the two analyses rely on overlapping nuclear data, we do not regard this as a fully independent determination. Therefore, we use \texttt{PRIMAT} as our benchmark and \texttt{PArthENoPE}~\cite{Gariazzo:2021iiu} as an alternative prescription. The two are compared in App.~\ref{app:bbn}. The global PDG fit on the $D/H$ ratio is~\cite{ParticleDataGroup:2026mpi}
\begin{equation}
Y_D^{\rm PDG}=(2.508\pm0.029)\times10^{-5},
\end{equation}
while for the homogeneous abundance \texttt{PRIMAT} predicts
\begin{equation}
Y_D^0\simeq(2.444\pm 0.037  )\times10^{-5}, \qquad (\texttt{PRIMAT}). 
\end{equation}
This lies below the observed value, a $1.4\sigma$ tension with the PDG fit that grows to $1.7\sigma$ against the Cooke et al. measurement, $10^5Y_D=2.527\pm0.030$~\cite{Cooke:2017cwo}, adopted in Refs.~\cite{Pitrou:2020etk,Pitrou:2021vqr,Launders:2026ciu}. Baryon isocurvature always shifts the mean abundance upward and can therefore alleviate this tension. Combining the observational and \texttt{PRIMAT} prediction uncertainties 
\footnote{The \texttt{PRIMAT} uncertainty is obtained by Monte Carlo propagation of the
nuclear-rate uncertainties in the post-LUNA reaction network. The dominant contributions to D/H arise from $D(p,\gamma)^3{\rm He}$, $D(d,n)^3{\rm He}$ and $D(d,p)t$.} 
at the $1\sigma$ level, the isocurvature amplitudes that reconcile the \texttt{PRIMAT} prediction with the PDG value are approximately
\begin{equation}
{0.12}\lesssim
\sqrt{\langle S_B^2\rangle}
\lesssim0.29 , \qquad (\texttt{PRIMAT}).
\label{eq:SB_range_primat}
\end{equation}
Above the upper endpoint, the presence of too strong isocurvature fluctuations overproduces deuterium. Since $S_B$ is conserved on the scales considered here, this range applies directly to the isocurvature generated by DW annihilation or a FOPT.

The \texttt{PArthENoPE}~\cite{Gariazzo:2021iiu} prediction lies substantially closer to the observed central value, 
\begin{equation}
Y_D^0\simeq(2.512 \pm 0.042)\times10^{-5}, \qquad (\texttt{PArthENoPE})
\end{equation}
and therefore does not produce any tension with the observational value. Hence, the baryon isocurvature is constrained only from above
\begin{equation}
\sqrt{\langle S_B^2\rangle}
\lesssim{0.19}, \qquad (\texttt{PArthENoPE}).
\label{eq:SB_range_parthenope}
\end{equation}

\textit{\textbf{Size of Fluctuations.}}
We now apply these results to the two aforementioned cosmological interpretations of the PTA signal. Generalizing Eq.~\eqref{eq:Sb_estimate} to track the fraction $f_\gamma$ of the injected energy that actually heats the photons, a source depositing a fraction $\alpha_S$ (Eq.~\eqref{eq:alpha}) of the total radiation density with contrast $\delta_S$ imprints
\begin{equation}
\label{eq:Sb_source}
S_B\simeq-\frac34\,\frac{f_\gamma\alpha_S}{\Omega_\gamma+f_\gamma\alpha_S}\,\delta_S,
\end{equation}
where $\Omega_\gamma$ denotes the relative photon energy density.  For the energy scales under consideration, the energy injected into SM particles (a fraction we call the visible branching fraction $f_{\rm vis}$) thermalizes with the plasma, so that $f_\gamma=\Omega_\gamma f_{\rm vis}$ and $S_B\simeq-\tfrac34\,\alpha_S\delta_S/(1+\alpha_S)$ taking $f_{\rm vis}= 1$. The source contrast $\delta_S$ is set by the spatial distribution of domain walls or nucleated bubbles. A DW network in the scaling regime contains on average one wall per Hubble patch, inducing large density perturbations on these smoothing scales. At the time of annihilation the vacuum (V) and wall-tension (DW) energies become comparable, and the two fluctuations are correlated,
producing an order-one density contrast, 
\begin{equation}
\label{eq:delta_DW_sqr}
\langle\delta_{\rm DW+V}^2\rangle\equiv c_\mathrm{DW}, \qquad c_\mathrm{DW}\simeq1\,.
\end{equation}
We estimate these size of these fluctuations in App.~\ref{app:sourcefluctuations} and find $c_\mathrm{DW}=1$, which we will use as our fiducial value for the remainder of this work. Assuming complete decay into Standard Model particles in thermal equilibrium ($f_{\rm vis} = 1$), Eq.~\eqref{eq:Sb_source} yields
\begin{equation}
\sqrt{\langle (S_B^{\rm DW})^2\rangle} \simeq \frac{3}{4} \frac{\alpha_{\rm DW +V}}{1 + \alpha_{\rm DW+V}} \sqrt{\langle\delta_{\rm DW+V}^2\rangle},
\end{equation}
where we used $\alpha_{\rm DW+V}\simeq\tfrac52\,\alpha_{\rm DW}$ as found in simulations~\cite{Kawasaki:2014sqa}. Requiring
$\sqrt{\langle S_B^2 \rangle} \le 0.29$, cf. Eq.~\eqref{eq:SB_range_primat}, then bounds the energy strength to $\alpha_{\rm DW} \lesssim 0.25$, partially ruling out the strong DW networks needed to fit the PTA amplitude, unless reheating is dark-dominated.
 
For FOPTs, the density fluctuations arise from the stochastic number of nucleation sites per Hubble volume (App.~\ref{app:sourcefluctuations})~\cite{Elor:2023xbz}, 
\begin{equation}
\sqrt{\langle \delta_{\rm FOPT}^2\rangle}\simeq 1.4d_bH_\star,\qquad d_b=(8\pi)^{1/3}/\beta,
\label{eq:deltaFOPT}
\end{equation}
where $d_b$ is the mean bubble separation at percolation~\cite{Hindmarsh:2019phv} and $\beta$ the transition rate. Slow transitions ($\beta/H_\star \sim \mathcal{O}(1$--$10)$) with large latent heat ($\alpha_{\rm FOPT} \gtrsim 1$) thus generate large density fluctuations on horizon scales, and in the ultra-slow limit ($d_b H_\star \sim 1$) Eq.~\eqref{eq:deltaFOPT} smoothly reproduces the order unity DW value ($c_{\rm DW} \sim 1$)~\footnote{Having $c_{\rm DW}\simeq 1$ in Eq.~\eqref{eq:delta_DW_sqr} instead of $c_{\rm DW}\simeq 1.4$ as in Eq.~\eqref{eq:deltaFOPT} is conservative.}.

For $f_{\rm vis}=1$, Eq.~\eqref{eq:Sb_source} gives
\begin{equation}
\sqrt{\langle (S_B^{\rm FOPT})^2 \rangle} \simeq \frac{3}{4} \left( \frac{\alpha_{\rm FOPT}}{1+\alpha_{\rm FOPT}} \right) \left( \frac{4H_\star}{\beta} \right) \, ,
\end{equation}
so that the upper bound of Eq.~\eqref{eq:SB_range_primat}, $\sqrt{\langle S_B^2 \rangle} \le 0.29$, translates into
\begin{equation}
\frac{\beta}{H_\star} \gtrsim 10.3 \left( \frac{\alpha_{\rm FOPT}}{1+\alpha_{\rm FOPT}} \right) \, .
\end{equation}
 
\begin{figure*}
    \centering
    \includegraphics[width=0.49\linewidth]{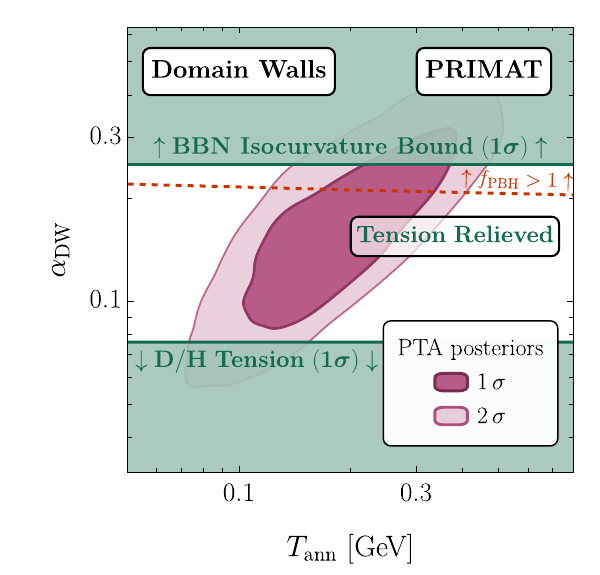}
    \includegraphics[width=0.49\linewidth]{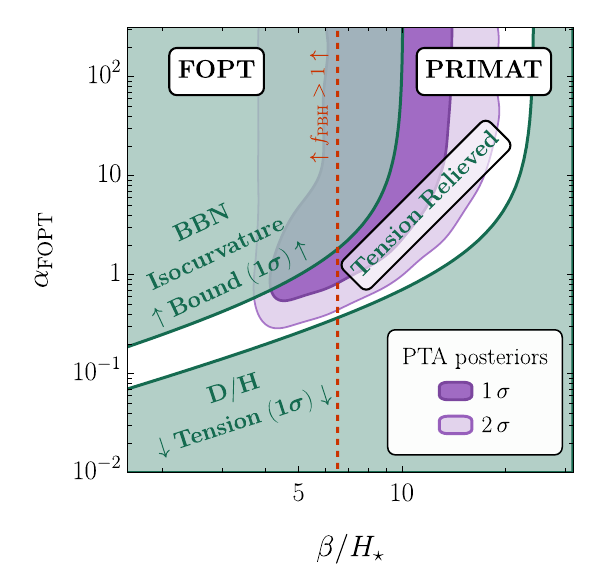}
\caption{\it Baryon-isocurvature constraint from BBN, overlaid on the PTA posteriors for the domain-wall (left) and first-order phase-transition (right) interpretations of the NANOGrav 15-year signal. The deuterium abundance is computed with \texttt{PRIMAT}, whose lower central value of $Y_D$ leaves a mild $D/H$ tension. The white regions, ${0.12}\lesssim\sqrt{\langle S_B^2\rangle}\lesssim0.29$, alleviate it, while the green regions either under- or overproduce deuterium. Filled contours show the $68\%$ ($1\sigma$, dark) and $95\%$ ($2\sigma$, light) posteriors obtained from the NG15 data~\cite{NANOGrav:2023gor} using the GW templates of~\cite{NANOGrav:2023hvm,Lewicki:2025hxg}. We assume $T_{\rm ann},T_{\rm reh}<3\,\mathrm{TeV}$, for which the isocurvature is conserved until BBN. \textbf{Left:} In the $(T_{\rm ann},\alpha_{\rm DW})$ plane, this yields temperature-independent bounds on $\alpha_{\rm DW}$, while leaving part of the PTA preferred region compatible with $D/H$. \textbf{Right:} In the $(\beta/H_\star,\alpha_{\rm FOPT})$ plane, the bound excludes strong, slow transitions, removing a substantial fraction of the PTA posterior. Dashed red lines indicate PBH overproduction, $f_{\rm PBH}>1$, for DWs~\cite{Gouttenoire:2023ftk,Gouttenoire:2023gbn,Gouttenoire:2025ofv,Ferreira:2024eru} and FOPTs~\cite{Kodama:1982sf,Liu:2021svg,Kawana:2022olo,Gouttenoire:2023naa,Baldes:2023rqv,Gouttenoire:2023pxh,Lewicki:2024ghw,Lewicki:2024sfw}, with the latter at $\beta/H_\star\simeq6.5$~\cite{Lewicki:2024ghw}.}
    \label{fig:Excl}
\end{figure*}
\textit{\textbf{Implications for PTA Interpretation.}} 
PTAs report evidence for a stochastic GW background at nanohertz frequencies~\cite{NANOGrav:2023gor,EPTA:2023fyk,Reardon:2023gzh,Xu:2023wog}, corresponding to horizon entry at $T\sim0.1$–$1~{\rm GeV}$, around the QCD epoch. To interpret the signal in terms of DWs or FOPTs, we fit the NANOGrav 15-year dataset~\cite{NANOGrav:2023gor} with state-of-the-art templates for the present-day GW spectrum $\Omega_{\rm GW}(f)$, parametrized by $(T_{\rm ann},\alpha_{\rm DW})$ for DWs~\cite{NANOGrav:2023hvm} and by $(T_{\rm reh},\alpha_{\rm FOPT},\beta/H_\star)$ for FOPTs~\cite{Lewicki:2025hxg}. Fig.~\ref{fig:Excl} compares the resulting posteriors with the BBN limit derived above. A comparison with alternative GW templates is given in App.~\ref{app:pta}.

The corresponding BBN exclusion follows from Eq.~\eqref{eq:Sb_source}, evaluated over the same parameter space. Since $S_B$ is conserved for the models considered here, the constraint depends only on the isocurvature amplitude and is independent of the annihilation or reheating temperature. The resulting BBN sensitivity persists up to source scales corresponding to the Hubble horizon at $\sim 3\ {\rm TeV}$~\cite{Bagherian:2025puf}, with modifications on scales below the neutron diffusion length that in fact enhance it~\footnote{Neutron diffusion does not erase the baryon inhomogeneity. For modes with $k\,d_n\gtrsim1$ but $k\,d_p\lesssim1$, the neutron fluctuations are smoothed while a proton inhomogeneity survives, so BBN remains sensitive to the surviving structure~\cite{Bagherian:2025puf}. We treat this regime conservatively and defer the details to App.~\ref{app:bbn}.}. The white regions in Fig.~\ref{fig:Excl} indicate where the induced isocurvature raises the predicted $D/H$ abundance toward the observed value, thereby alleviating the deuterium tension for both DWs and FOPTs. Below these regions, the isocurvature amplitude is too small to account for the discrepancy, denoted as $D/H$ tension. 
At larger source amplitudes, the same effect overproduces deuterium, yielding strong exclusions that we denote as the BBN isocurvature bound. For strong FOPTs, $\alpha_{\rm FOPT}\gg1$, the BBN isocurvature constraint provides the strongest current bound over the relevant parameter space, while for DWs it excludes $\alpha_{\rm DW}\gtrsim0.25$. The latter is competitive with the existing constraint from PBH overproduction, $f_{\rm PBH}>1$, shown in red. The PBH bounds should, however, be interpreted with some caution, as they rely on exponentially rare high-density peaks and are consequently sensitive to the assumed statistics of the perturbations. For FOPTs, gauge ambiguities in the cosmological perturbation theory underlying PBH formation may further weaken, or potentially invalidate, this constraint~\cite{Flores:2024lng,Franciolini:2025ztf,Wang:2026zvz}. {The same fluctuations also seed ultra-compact minihalos, which exclude WIMP dark matter in these scenarios~\cite{Liu:2023hte,Gouttenoire:2025wxc}.} By contrast, the BBN bound derived here probes the typical source fluctuations directly{, with no assumption about dark matter}. A direct probe of the source fluctuations is also possible with CMB spectal distortions, which are however more sensitive at lower frequencies~\cite{Ramberg:2022irf}.

For both DW and FOPTs, part of the $1\sigma$ PTA preferred region overlaps with the parameter space that alleviates the $D/H$ tension. DW annihilation and FOPTs can therefore simultaneously account for the nHz SGWB and shift the BBN prediction toward the observed deuterium abundance, providing an additional consistency test of these interpretations.

Fig.~\ref{fig:parthenope} compares the \texttt{PRIMAT} results discussed above with those obtained using \texttt{PArthENoPE}. In the latter case the predicted and observed central values are aligned, so no $D/H$ tension arises and the isocurvature is bounded only from above. The constraint is correspondingly stronger, and rules out the majority of the PTA preferred parameter space.

The DW interpretation is less strongly constrained than the FOPT scenario and can alleviate the deuterium tension over a larger fraction of the PTA preferred parameter space. This difference reflects the efficiency of GW production. A DW network in the scaling regime sources GWs on scales of order the Hubble radius, whereas FOPT sources are subhorizon. The latter therefore require larger density fluctuations to produce the same GW amplitude and consequently reach the deuterium-overproduction threshold more rapidly.

\begin{figure*}
    \centering
    \includegraphics[width=0.49\linewidth]{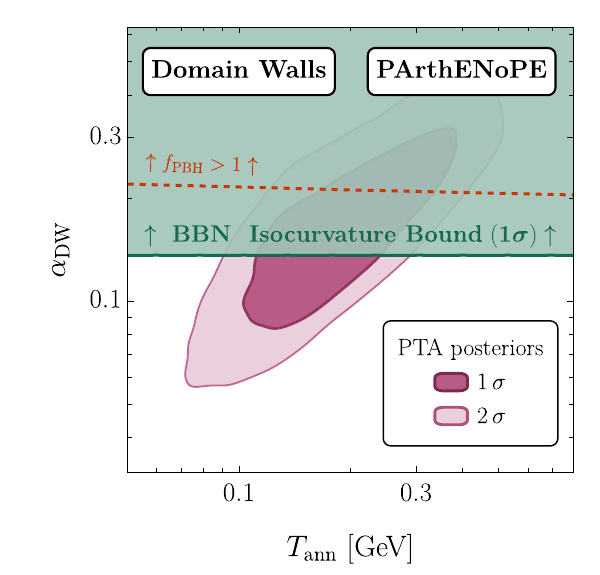}
    \includegraphics[width=0.49\linewidth]{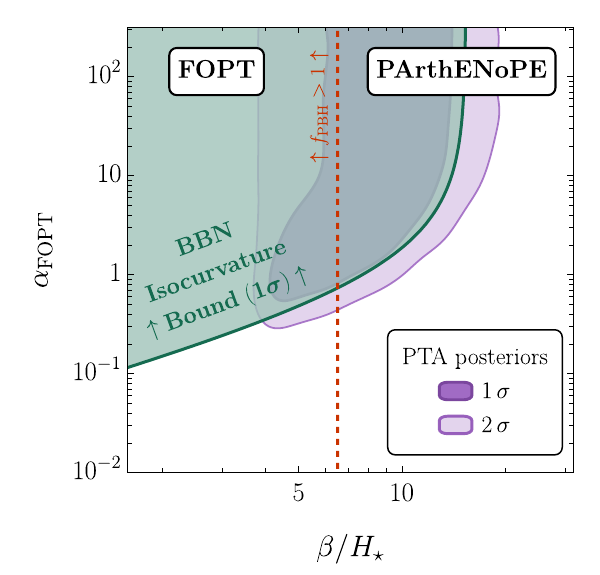}
\caption{\it Same as Fig.~\ref{fig:Excl}, but with the deuterium abundance computed using \texttt{PArthENoPE}. Its central value is aligned with the observed one, so no $D/H$ tension arises and the isocurvature is constrained only from above, $\sqrt{\langle S_B^2\rangle}\lesssim{0.19}$, by deuterium overproduction. The resulting exclusion green regions ruled out the majority of PTA preferred regions for both DWs and FOPTs.}
    \label{fig:parthenope}
\end{figure*}

So far we have assumed that a sizeable fraction of the released energy reheats the SM. If reheating instead proceeds dominantly into a dark sector ($f_{\rm vis}\to0$), the photon bath is barely perturbed, the baryon isocurvature vanishes ($S_B\propto f_{\rm vis}\to0$), and the BBN bound disappears. Instead, stable, relativistic dark particles are produced and survive as dark radiation and contribute to the effective number of relativistic species, 
\begin{equation}
8.6 \left(1 - f_{\rm vis}\right) \alpha_S \left( \frac{g_{\star s}(T_{\rm dec})}{g_{\star s}(T_\star)} \right)^{1/3}\lesssim\Delta N_{\rm eff},
\end{equation}
with $\Delta N_{\rm eff}\lesssim(0.07,\,0.17)$ at $(1\sigma,2\sigma)$ from Planck\,+\,ACT~\cite{Planck:2018vyg,AtacamaCosmologyTelescope:2025nti}. The baryon isocurvature and
$\Delta N_{\rm eff}$ observables therefore scale oppositely with the visible branching, $S_B\propto f_{\rm vis}$ and $\Delta N_{\rm eff}\propto(1-f_{\rm vis})$. Together, they cover the full range of branching fractions (for details see App.~\ref{app:DR}).

\textit{\textbf{Discussion and conclusion.}}
We have shown that the same inhomogeneities that allow DWs or FOPTs to source a nanohertz GW background also imprint a baryon isocurvature perturbation on the primordial plasma. This perturbation is conserved from the source epoch down to BBN, where the nonlinear dependence of the light-element abundances on the local baryon-to-photon ratio turns it into an observable shift in the primordial deuterium abundance $D/H$. BBN therefore constrains the spatial structure of the source, rather than only its mean energy density, directly linking the large inhomogeneities required for efficient GW production to an independent cosmological observable. Remarkably, with \texttt{PRIMAT} bound ${0.12}\lesssim\sqrt{\langle S_B^2\rangle}\lesssim{0.29}$, this shift can alleviate the existing $D/H$ tension while simultaneously accounting for the PTA signal, strengthening the case for a common BSM origin of both observations. At larger isocurvature amplitudes, however, deuterium is overproduced. Sufficiently strong FOPTs and abundant DW networks rapidly exceed this range, yielding constraints competitive with existing bounds for DWs and, for FOPTs, stronger than current limits in the literature. As a second benchmark we use \texttt{PArthENoPE}, for which the primordial $D/H$ determination and the standard BBN prediction share the same central value. The allowed isocurvature window then shrinks substantially, excluding most of the DW and FOPT parameter space ($\sqrt{\langle S_B^2\rangle}\lesssim {0.19}$) capable of explaining the PTA signal. This illustrates the power of primordial abundances as probes of BSM physics, especially in regard to future improvements in deuterium-abundance predictions and observations.

Moreover, this probe is complementary to existing cosmological limits. Whereas $\Delta N_{\rm eff}$ traces energy hidden in a dark sector, baryon isocurvature probes the inhomogeneous reheating of the visible plasma. The two constrain opposite ends of the source branching fraction, together covering a broad range of reheating scenarios~\cite{Bagherian:2025puf}. It is evaded only if baryogenesis follows the same spatial pattern as the injected photons, keeping $\eta_B$ fixed. {Finally, the moving walls that source the GW signal, during either DW annihilation or FOPT, also induce bulk motion of the plasma. The resulting turbulence could cascade down to the nucleon diffusion scale, and we leave its impact on our findings to future work.}

\textit{\textbf{Acknowledgments.}}  
The authors thank Majid Ekhterachian, Julien Froustey and Cyril Pitrou for valuable discussions. The authors acknowledge support by the Cluster of Excellence ``PRISMA$^{++}$'' funded by the German Research Foundation (DFG) within the German Excellence Strategy (Project No. 390831469). NL is grateful to the German Academic Scholarship Foundation for the award of a PhD fellowship.

\bibliographystyle{apsrev4-1}
\bibliography{Letter}
\clearpage

\appendix
\onecolumngrid
\fontsize{11}{13}\selectfont

\makeatletter
\long\def\@makecaption#1#2{%
  \vskip\abovecaptionskip
  \parbox{\hsize}{%
    \fontsize{10.5}{12.5}\selectfont 
    \justifying
    #1.\ #2%
  }%
  \vskip\belowcaptionskip
}
\makeatother

\clearpage
\vspace{1cm}

\begin{center}
\textbf{\large Supplemental Material for ``Big Bang Nucleosynthesis confronts Domain Walls and First-Order Phase Transitions as the Pulsar Timing Array Explanations''}
\end{center}

\noindent
\renewcommand{\tocname}{\it
\vspace{-1cm}
Fatemeh Elahi, Diego Fortini, Yann Gouttenoire, Nicholas Leister, Pedro Schwaller
}

\titleformat{\section}
{\normalfont\fontsize{12}{14}\bfseries  \centering }{\thesection.}{1em}{}
\titleformat{\subsection}
{\normalfont\fontsize{12}{14}\bfseries \centering}{\thesubsection.}{1em}{}
\titleformat{\subsubsection}
{\normalfont\fontsize{12}{14}\bfseries \centering}{\thesubsubsection)}{1em}{}

\titleformat{\paragraph}
{\normalfont\fontsize{12}{14}\bfseries  }{\thesection:}{1em}{}

 {
 \hypersetup{linkcolor=zima_blue}
 \tableofcontents
 }

 \clearpage
\section{Initial Perturbations}
\label{app:in_pert}
\subsection{Photon perturbation from inhomogeneous heating}
\noindent
We now compute the photon perturbation $\delta_\gamma$ produced when an
inhomogeneous source injects energy into the photons, i.e. the estimate used in
Eq.~\eqref{eq:Sb_estimate}.

Before the injection the photons are smooth and all the inhomogeneity is in the
source. Two numbers describe the source: its energy relative to the radiation
bath, $\alpha_S\equiv\rho_S/\rho_{\rm SM}$, and its density contrast
$\delta_S=\delta\rho_S/\rho$. At the injection a fraction $f_\gamma$ of the source energy is transferred to the photons. The source has a mean energy density $\alpha_S\rho_{\rm SM}$ and fluctuations of the size
$\alpha_S\delta_S\rho_{\rm SM}$. The heating transfers
$f_\gamma\alpha_S\rho_{\rm SM}$ to the mean photon density and
$f_\gamma\alpha_S\delta_S\rho_{\rm SM}$ to its perturbation. The photon
contrast is the ratio of the two,
\begin{equation}
\label{eq:dgamma_app}
\delta_\gamma=\frac{f_\gamma\alpha_S}{\Omega_\gamma+f_\gamma\alpha_S}\,\delta_S,
\qquad \Omega_\gamma\equiv\frac{\rho_\gamma}{\rho_{\rm SM}}.
\end{equation}
The photon contrast is just the source contrast scaled by the fraction of the photon energy that came from the source. In thermal equilibrium and assuming radiation domination, the background quantities and fluctuations only depend on temperature 
\begin{equation}
    \delta\rho\propto4T^3\delta T
\end{equation}
and are distributed among all species in thermal contact. This yields locally and globally 
\begin{equation}
\label{eq:delta_equilibrium}
    \frac{\delta \rho_a}{\delta \rho_b}= \frac{\rho_a}{\rho_b} =\frac{g_{\star,a}(T)}{g_{\star,b}(T)}
\end{equation}
and hence $f_\gamma=f_{\rm vis}\Omega_\gamma$.
\section{Source fluctuations}
\label{app:sourcefluctuations}
{\noindent Here we estimate the density contrast $\delta_S$ carried by the two sources considered in the main text, together with the comoving scale on which it is imprinted.}

\subsection{Domain Wall Network}
\noindent
Domain walls form when a discrete symmetry is spontaneously broken. Causally disconnected patches settle into different degenerate vacua, and the two-dimensional boundaries between them survive as topological defects~\cite{Kibble:1976sj}. A minimal realization is a real scalar $\phi$ with a $\mathbb{Z}_2$-symmetric double well,
\begin{equation}
    V(\phi)=\tfrac{\lambda}{4}(\phi^2-v^2)^2,
\end{equation}
whose minima $\phi=\pm v$ are separated
by walls of surface tension $\sigma_{\rm DW}\sim\sqrt{\lambda}\,v^3$. Shortly after formation the network reaches the so-called scaling regime by the recombination of nearby walls bounded by causality. In this regime, one DW per Hubble patch survives on average, which gives the characteristic correlation length $L(t)\sim H^{-1}$. Simulations find $\rho_{\rm DW}=\mathcal A\sigma_{\rm DW}/t$ with $\mathcal A\simeq0.8$~\cite{Hiramatsu:2013qaa}. An exact $\mathbb{Z}_2$ is cosmologically excluded, since such walls would dominate the universe eventually by red-shifting slower than the radiation component.  A small bias $\Delta V$ between the two minima prevents this by driving the true-vacuum regions to expand until the network annihilates
\cite{Kawasaki:2014sqa}. This happens, once the vacuum pressure $\mathcal P_V =\Delta V$ overcomes the wall tension pressure $\mathcal{P}_T\simeq C_d\rho_{\rm DW}$, with $C_d$ being a dimensionless $\mathcal{O}(1)$ DW model parameter.  Equating the two yields:
\begin{equation}
\label{eq:sigmaDW_App}
    \sigma_{\rm DW}
    =
    \left(
    \frac{\pi^2 g_\star (T_{\rm ann})}{40\,\mathcal{A}^2}
    \right)^{1/2}
    \alpha_{\rm DW}\,
    M_{\rm Pl}\,
    T_{\rm ann}^2 .
\end{equation}
Fluctuations in both the vacuum and the DW components peak at the correlation scale. Since these will be imprinted on the SM bath, we average over the Hubble size at annihilation, when bias and wall tension contributions are similar in magnitude. We denote the spatial average by $\langle\cdots\rangle$ and denote
\begin{equation}
\delta_i(\mathbf{x})
=
\frac{\rho_i(\mathbf{x})-\bar{\rho}_i}{\bar{\rho}_i},
\qquad
\langle \rho_i^{n} \rangle
=
\frac{1}{\mathcal{V}}
\int_{\mathcal{V}}
\mathrm{d}^3x\,
\rho_i^n(\mathbf{x}),
\end{equation}
where the smoothing volume $\mathcal{V}$ is chosen to be the Hubble volume at annihilation.
The network perturbations are then described by the variance $\langle\delta_{\rm DW}^2\rangle$.

{\bf\emph{Vacuum contribution.}} The false-vacuum domains occupy a fraction $p$ of a given volume and carry an energy density $\Delta V$. The mean density contrast then vanishes, while its variance does not.
\begin{equation}
\langle\delta_{\rm V}\rangle=0,\qquad
\langle\delta_{\rm V}^2\rangle=\frac{1-p}{p},
\end{equation}
This is order unity in the scaling regime with $p\simeq\tfrac12$.

{\bf \emph{Wall contribution.}} Using
$\rho_{\rm DW}=\sigma_{\rm DW}/L$ and $L\simeq t/\mathcal A$, the enclosed area $A$ varies by an order-one fraction between regions on the (horizon sized) smoothing scale,
\begin{equation}
\langle\delta_{\rm DW}^2\rangle
=\frac{\langle A^2\rangle-\langle A\rangle^2}{\langle A\rangle^2}\simeq 1 .
\end{equation}
This stays finite in the thin-wall limit, since it depends only on the integrated wall energy $\sigma_{\rm DW}L^2$.

{\bf\emph{Combined.}} With mean energy fractions $f_{\rm V}=\bar\rho_{\rm V}/\bar\rho$ and
$f_{\rm DW}=\bar\rho_{\rm DW}/\bar\rho$, the total contrast is
$\delta_{\rm DW+V}=f_{\rm V}\delta_{\rm V}+f_{\rm DW}\delta_{\rm DW}$. Because a wall is
the boundary of a false-vacuum domain, a smoothing cell containing more false
vacuum also encloses more wall area. At the scaling scale $L$ both normalised
fluctuations trace the same domain pattern, and to the order-one accuracy of this
estimate we take them equal, $\delta_{\rm DW}=\delta_{\rm V}$. The cross term then
adds coherently,
\begin{equation}
\langle\delta_{\rm DW+V}^2\rangle=f_{\rm V}^2+f_{\rm DW}^2+2f_{\rm V}f_{\rm DW}
=(f_{\rm V}+f_{\rm DW})^2=1 .
\end{equation}
Hence, independently of how the energy is shared between walls and vacuum, the
density contrast smoothed over the network scale is order unity,
\begin{equation}
\langle\delta_{\rm DW+V}^2\rangle\simeq 1.
\end{equation}
A biased $\mathbb{Z}_2$ wall network therefore realises the order-one contrast
$\delta_{\rm DW}\simeq\mathcal{O}(1)$ used in the main text to source the baryon
isocurvature.

\subsection{First-Order Phase Transition}
\noindent
A first-order phase transition occurs when a scalar field is trapped in a metastable minimum separated from the true vacuum by a potential barrier. The transition proceeds by nucleation of true-vacuum bubbles, which expand, collide and convert the latent heat released into radiation, reheating the bath. The nucleation rate per unit volume grows exponentially near completion,
\begin{equation}
\Gamma(t)\simeq\Gamma_\star\,e^{\beta(t-t_\star)},
\end{equation}
so the transition completes within a time $\sim\beta^{-1}$. The ratio $\beta/H_\star$
measures how rapidly it proceeds, while the vacuum energy released relative to the
radiation bath defines the strength $\alpha_{\rm FOPT}$.

Because bubble nucleation is stochastic and only a finite number of bubbles form per Hubble volume, different super-horizon patches complete the transition at slightly different times $t_\star$~\cite{Elor:2023xbz}. The released radiation redshifts as $\rho\propto a^{-4}$, so a fluctuation in the completion time enters only through the scale factor at completion. Evaluating at a  to a common later slice, yields
\begin{equation}
\delta_{\rm FOPT}=\frac{\delta\rho}{\bar\rho}
=4\,\delta\ln a(t_\star)
=4\,\frac{\dot a}{a}\bigg|_\star\delta t_\star
=4\,H_\star\,\delta t_\star.
\end{equation}
The completion-time fluctuation is set by the finite bubble statistics. It is of order the transition duration, and peaks at $\beta^2\langle\delta t_\star^2\rangle\simeq1$~\cite{Elor:2023xbz}, giving:
\begin{equation}
\sqrt{\langle\delta_{\rm FOPT}^2\rangle}\simeq 4\,\frac{H_\star}{\beta}.
\end{equation}
A fast transition ($\beta/H_\star\gg1$) is therefore nearly homogeneous, whereas the strong, slow transitions that dominate the PTA signal carry the largest inhomogeneity, as used in the main text. 

\subsection{Source Scales}
\noindent
Both DWs and FOPTs source their fluctuations on roughly horizon scales. For a given annihilation or transition temperature $T_\star$, the comoving Hubble scale is
\begin{equation}
k_H(T_\star )\equiv a_\star H_\star \simeq 1.7\times10^7\,{\rm Mpc}^{-1} \left(\frac{T_\star}{1\,{\rm GeV}}\right) \left(\frac{g(T_\star)}{106.75}\right)^{1/2} \left(\frac{g_{\star s}(T_\star)}{106.75}\right)^{-1/3}.
\label{eq:kHdiff}
\end{equation}
In the scaling regime a DW network contains on average one wall per Hubble patch, so its inhomogeneity is set by the horizon scale,
\begin{equation}
k_{\rm DW}\sim k_H(T_{\rm ann}),
\label{eq:kDWdiff}
\end{equation}
whereas a FOPT completes through the nucleation of many bubbles per Hubble volume, whose mean separation at collision $R_\star\simeq v_w/\beta$ (with $v_w$ the bubble-wall velocity) sets the relevant scale,
\begin{equation}
k_{\rm FOPT}\simeq \frac{a_\star \beta}{v_w} = \frac{\beta}{H_\star }\,\frac{k_H(T_\star )}{v_w}.
\label{eq:kbubblediff}
\end{equation}
\clearpage
\section{Cosmological perturbation theory}
\noindent
\label{app:perturbation}
In this appendix we derive the equations needed to track the evolution of the baryon-photon isocurvature. On large scales, the Universe is well approximated by a homogeneous, isotropic, and spatially flat Friedmann–Lemaître–Robertson–Walker (FLRW) spacetime. In conformal time $\eta$, the line element reads
\begin{equation}
\label{eq:FLRW}
\mathrm{d}s^2
= a^2(\eta)\left(-\mathrm{d}\eta^2 + \delta_{ij}\,\mathrm{d}x^i \mathrm{d}x^j\right).
\end{equation} 
For a multiple perfect fluid with total energy density $\rho$ and equation-of-state parameter $\omega$, Einstein equations reduce to the standard Friedmann and continuity equations
\begin{align}
\label{eq:Hubble_prime}
\mathcal{H}^2 \equiv\left(\frac{a'}{a}\right)^2 = \frac{8\pi G}{3}\overline{\rho}a^2, \qquad
\mathcal{H}' = -\frac{1+3\omega}{2}\mathcal{H}^2,\qquad
\overline{\rho}' + 3(1+\omega)\mathcal{H}\overline{\rho} = 0.
\end{align}
Here, primes denote derivatives with respect to the conformal time $\eta$. Relative energy density fractions for each species $a$ are denoted by $\Omega_a\equiv\bar\rho_a/\bar\rho$ where the total average energy density $\bar\rho$ and the total equation of state parameter $\omega$ read~\cite{Peter:2013avv}
\begin{equation}
    \bar \rho = \sum_a \bar \rho_a, \qquad  \omega = \sum_a\Omega_a\omega_a.\label{eq:omega_cs_multifluid}
\end{equation}

\subsection{Perturbation Equations}
\noindent
At linear order the perturbed FLRW metric in Newtonian gauge becomes
\begin{equation}
ds^2=a^2[-(1+2\Psi)d\eta^2+(1-2\Phi)\delta_{ij}dx^idx^j].
\end{equation}
Different particle species $a$ 
enter through perturbations of the total energy-momentum tensor $T^{\mu}{}_{\nu}\equiv\sum_a T^{\mu}{}_{\nu}^{(a)}$~\cite{Kodama:1985bj,Ma:1995ey},
\begin{align}
T^{\mu}{}_{\nu} &=
\begin{pmatrix}
-(\bar\rho+\delta\rho) & (\bar\rho+\bar P)\,v_j
\\[6pt]
-(\bar\rho+\bar P)\,v^i & (\bar P+\delta P)\,\delta^{i}{}_{j}+\Pi^{i}{}_{j}
\end{pmatrix},
\end{align}
with the total energy density contrast $\delta\equiv\delta\rho/\overline{\rho}$, pressure perturbation $\delta P$, bulk velocity $v_i$ ($\hat{\partial}_i v\equiv v_i= i\hat{k}v$) and  anisotropic stress $\Pi$
\begin{equation}
   \delta=\sum_a\Omega_a\delta_a, \qquad \delta  P = \sum_a \delta P_a, \qquad (1+\omega)v=\sum_a\Omega_a(1+\omega_a)v_a, \qquad (1+\omega)\Pi=\sum_a\Omega_a(1+\omega_a)\Pi_a.
\end{equation}
 The total $T^\mu{}_\nu$ sources
$\Phi,\Psi$ through the Einstein equations. At linear order and in Fourier space with wavevector $k$ the Einstein equations become~\footnote{We use the convention of Ref. ~\cite{Baumann:2022mni}, which relates to Ref.~\cite{Ma:1995ey} by $v_{\text{BAU}}=-\theta_{\text{MB}}/k$, $\theta_{\text{MB}}\equiv iv_{\text{MB}}^ik_i$, $\Pi_{\text{BAU}}^{(a)}= \frac32 \sigma_a^{\text{MB}}$, $\Pi_{\text{MB}}=\frac{\bar \rho + \bar P}{\bar P}\Pi_\text{BAU}$.}
\begin{align}
-k^2 \Phi - 3\mathcal H\big(\Phi' + \mathcal H \Psi\big) &= \frac{3}{2}\mathcal{H}^2\,\delta, \label{eq:E00}\\
\Phi' + \mathcal H \Psi &= -\frac{3}{2}\mathcal{H}^2(1+\omega)\, \frac{v}{k}, \label{eq:E0i}\\
k^2(\Phi-\Psi) &= 3\mathcal{H}^2(1+\omega)\,\Pi, \label{eq:Estress}\\
\Phi'' + \mathcal{H}\Psi' + 2\mathcal{H}\Phi'
- \frac{1}{3}k^2(\Psi - \Phi)
+ (2\mathcal{H}' + \mathcal{H}^2)\Psi
&= \frac{3}{2}\mathcal{H}^2\frac{\delta P}{\overline{\rho}}. \label{eq:Eii}
\end{align}
with $\Pi_{ij}\equiv-(\overline{\rho} + \overline{P})\hat{k}_{\langle i}\hat{k}_{j \rangle}\Pi$, where $\hat{k}_{\langle i}\hat{k}_{j\rangle}
\equiv
\hat{k}_i\hat{k}_j
-\frac{1}{3}\delta_{ij}$.
We furthermore introduce
\begin{equation}
   \delta_{\rm nad} \equiv \frac{\delta P_{\rm nad}}{\overline{\rho}},\qquad \textrm{with}\quad \delta P_{\rm nad} \equiv \delta P - c_s^2 \delta \rho,\qquad \textrm{and}\qquad  c_s^2 \equiv \frac{\overline{P}'}{\overline{\rho}'}.
   \label{eq:delta_nad_definition}
\end{equation}
For the mixture of fluids the non-adiabatic pressure perturbation can be expressed as
\begin{equation}
 \delta_{\rm nad}=\sum_a \Omega_a\delta_{\rm nad,a}-\sum_{a}(c_s^2-c_a^2)\Omega_a \delta_a,\qquad \textrm{where}\quad \delta_{\rm nad,a}\equiv \frac{\delta P_a}{\rho_a} -c_a^2\frac{\delta\rho_a}{\rho_a}.\label{eq:delta_p_multifluid}
\end{equation}
Assuming that the fluids are all barotropic, meaning that their pressure is completely determined by the energy density, implying $\delta P_{\rm nad,a}=0$.
Eq.~\eqref{eq:delta_p_multifluid} then becomes
\begin{equation}
 \delta_{\rm nad}=\sum_{a}(c_a^2-c_s^2)\Omega_a \delta_a. \label{eq:delta_p_multifluid_2}
\end{equation}
During radiation domination the non-adibatic density contrast vanishes $\delta_{\rm nad} \approx 0$ since all dominant species have the same equation of state parameter $\omega_a=1/3$. Combining Eqs.~\eqref{eq:E00},~\eqref{eq:Estress} and~\eqref{eq:Eii} yields: 
\begin{equation}
\Phi'' + 4\mathcal{H}\,\Phi' + \frac{1}{3}k^2\Phi
= -\frac{24\mathcal{H}^4\Pi}{3k^2} - \frac{4}{3}\mathcal{H}^2\Pi + \frac{4\mathcal{H}^3\Pi'}{k^2}\,.
\label{eq:Einstein_combined_approx_df}
\end{equation}
The system is closed by covariant energy-momentum conservation, $\nabla^\mu T_{\mu\nu}=0$, which gives the continuity and Euler equations for each fluid component $a$:
\begin{align}
    \delta_a'&=(1+\omega_a)\left(kv_a+3\Phi'\right),\\
    v_a'&=-\mathcal{H}(1-3\omega_a)v_a-\frac{\omega_a}{1+\omega_a}k\delta_a+\frac23 k\Pi_a-k\Psi-\sum_b\Gamma_{ab}(v_a-v_b).
    \label{eq:delta_v_general}
\end{align}
Enthalpy conservation moreover implies 
\begin{equation}
     h_a\Gamma_{ab}=h_b\Gamma_{ba}\qquad {\rm with}\qquad h_a\equiv \rho_a+ P_a.
\end{equation}

\subsection{Full Fluid System}
\noindent
We are interest in the period before BBN. At that time, the universe is dominated by electrons, baryons, photons and neutrinos. In the case of baryons, there are thermal and non-thermal contributions to the number density. We therefore split the baryon into its symmetric $n_{B_+}\equiv n_B+n_{\bar B}$ and anti-symmetric component $n_{B_-}\equiv n_B-n_{\bar B}$ and assume, for simplicity, that 
\begin{equation}
    n_{B_+}\approx (2\bar n_B^{\rm thermal} +\eta_B \bar n_\gamma), \qquad  n_{B_-}=  \eta_B \bar n_\gamma.
\end{equation}
Introducing the corresponding perturbations 
\begin{equation}
\delta_{B_+}
=
\frac{\delta n_B+\delta n_{\bar B}}
{\bar n_B+\bar n_{\bar B}},
\qquad
\delta_{B_-}
=
\frac{\delta n_B-\delta n_{\bar B}}
{\bar n_B-\bar n_{\bar B}}
\end{equation}
and assuming that baryons and antibaryons share the same velocity $v_B\simeq v_{\bar B}$ yields
the closed fluid system
\begin{equation}
\label{eq:full_system}
\begin{aligned}
\delta_\gamma' &= \tfrac43 k v_\gamma+4\Phi',\\
v_\gamma' &= -\tfrac14 k\delta_\gamma+\tfrac23 k\Pi_\gamma-k\Psi
-\Gamma_{\gamma e}(v_\gamma-v_e)-\Gamma_{\gamma B}(v_\gamma-v_B),\\
\delta_e' &= \tfrac43 k v_e+4\Phi',\\
v_e' &= -\tfrac14 k\delta_e+\tfrac23 k\Pi_e-k\Psi
-\Gamma_{e\gamma}(v_e-v_\gamma)-\Gamma_{eB}(v_e-v_B)-\Gamma_{e\nu}(v_e-v_\nu),\\
\delta_\nu' &= \tfrac43 k v_\nu+4\Phi',\\
v_\nu' &= -\tfrac14 k\delta_\nu+\tfrac23 k\Pi_\nu-k\Psi
-\Gamma_{\nu e}(v_\nu-v_e)-\Gamma_{\nu B}(v_\nu-v_B),\\
\delta_{B_-}' &= k v_B+3\Phi',\\
\delta_{B_+}' &= k v_B+3\Phi'+ \mathcal{S}_{B\bar B}^{\rm ann},\\
v_B' &= -\mathcal{H}v_B-k\Psi
-\Gamma_{Be}(v_B-v_e)-\Gamma_{B\gamma}(v_B-v_\gamma)-\Gamma_{B\nu}(v_B-v_\nu),\\
\Phi'' + 4\mathcal{H}\,\Phi' + \frac{1}{3}k^2\Phi
&= -\frac{8\mathcal{H}^4\Pi}{k^2} - \frac{4}{3}\mathcal{H}^2\Pi + \frac{4\mathcal{H}^3\Pi'}{k^2}.
\end{aligned}
\end{equation}
Here we assumed that the baryon-anti-baryon annihilation does not produce any net baryon number. The source term $\mathcal S_{B\bar B}^{\rm ann}$ can be computed from the perturbed annihilation Eq.~\cite{Kolb:1990vq} 
\begin{equation}
\overline{n}_{B+}^\prime+3\mathcal H n_{B+}
 =
 -2a\langle\sigma v\rangle_{B\bar B}^{\rm ann}
 \left[
 \bar n_B\bar n_{\bar B}
 -
 n_B^{\rm eq}n_{\bar B}^{\rm eq}
 \right].
\end{equation} 
and reads
\begin{align}
\mathcal S_{B\bar B}^{\rm ann}
={}&
-\frac{2a\langle\sigma v\rangle_{B\bar B}^{\rm ann}}
{\bar n_{b+}}
\left[
\delta(n_B n_{\bar B})
-
\delta(n_B^{\rm eq}n_{\bar B}^{\rm eq})
-
\left(
\bar n_B\bar n_{\bar B}
-
n_B^{\rm eq}n_{\bar B}^{\rm eq}
\right)
\delta_{b+}
\right.
\nonumber\\
&\left.
\quad
+
\left(
\bar n_B\bar n_{\bar B}
-
n_B^{\rm eq}n_{\bar B}^{\rm eq}
\right)
\Psi
+
\left(
\bar n_B\bar n_{\bar B}
-
n_B^{\rm eq}n_{\bar B}^{\rm eq}
\right)
\delta_\lambda
\right].
\end{align}
where
\begin{equation}
 \delta(n_Bn_{\bar B})
 =
 \frac12
 \left[
 \bar n_{b+}^2\delta_{b+}
 -
 \bar n_{b-}^2\delta_{b-}
 \right],\quad  \delta(n_B^{\rm eq}n_{\bar B}^{\rm eq})
 =
 n_B^{\rm eq}n_{\bar B}^{\rm eq}
 \frac{d\ln(n_B^{\rm eq}n_{\bar B}^{\rm eq})}{d\ln T}
 \frac{\delta_\gamma}{4}, \quad  \delta_\lambda
 =
 \frac{d\ln\langle\sigma v\rangle_{B\bar B}^{\rm ann}}{d\ln T}
 \frac{\delta_\gamma}{4}.
\end{equation}
Reactions happen per proper time, but we evolve in conformal time. Since $d\tau_{\rm prop}=a(1+\Psi)d\eta$, the collision term gets the term proportional to $\Psi$.
Furthermore, $\delta _\lambda$ comes from the temperature dependence of $\langle\sigma v\rangle_{B\bar B}^{\rm ann}$. The two baryon isocurvature perturbations therefore are
\begin{equation}
 S_{B_-}
 =
 \delta_{B_-}-\frac34\delta_\gamma,
 \qquad
 S_{B_+}
 =
 \delta_{B_+}-\frac34\delta_\gamma.
\end{equation}
Their evolution is
\begin{equation}
 \frac{dS_{B_-}}{d\eta}
 =
k( v_B-v_\gamma),
 \qquad
 \frac{dS_{B_+}}{d\eta}
 =
 k(v_B-v_\gamma)+\mathcal S_{B\bar B}^{\rm ann}.
 \label{eq:S_B_evolution}
\end{equation}
Thus $S_{B_-}$ is conserved whenever $v_B\simeq v_\gamma$, whereas $S_{B_+}$ is not conserved during annihilation. At much later times, however, the two components become equivalent and therefore approach the same value, as we show explicitly in the next section. The BBN constraint applies in this late-time regime, so it suffices to compute the net-baryon isocurvature $S_{B_-}$.

It is also worth noting that the isocurvature $S_{B_-}=\delta_{B_-}-\tfrac34\delta_\gamma$ is gauge invariant. It is the standard entropy perturbation between baryons and photons~\cite{Kodama:1985bj}, and at first order it equals the fluctuations in the baryon-to-photon entropy ratio $\delta(n_B/s_\gamma)/(n_B/s_\gamma)$. Because it does not depend on the space time slicing, the conservation law~\eqref{eq:Sb_evolution} takes the same form in every gauge.  The bound on $S_B$ is physical and is not a gauge artifact.

\subsection{Reduced Fluid System}
\noindent
The electron scattering rates dominate over the remaining scales,
\begin{equation}
\Gamma_e\gg\Gamma_{i\neq e},\ \mathcal{H},
\end{equation}
electrons and baryons share a common velocity up to a small slip 
\begin{equation}
v_e-v_B=\mathcal{O}(\Gamma_e^{-1}),\qquad v_e\simeq v_B.
\end{equation}
Combing the electron and baryon Euler equation and setting $v_e\simeq v_b$ afterwards, one obtains
\begin{equation}
v_B'=-k\Psi
-\frac{h_B}{h_e+h_B}\,\mathcal{H}v_B
-\frac{h_e}{h_e+h_B}\,\tfrac14 k\delta_\gamma
+\frac{h_\gamma\,(\Gamma_{\gamma e}+\Gamma_{\gamma B})}{h_e+h_B}(v_\gamma-v_B)
+\frac{h_\nu\,(\Gamma_{\nu e}+\Gamma_{\nu B})}{h_e+h_B}(v_\nu-v_B)
\end{equation}
Here, we choose $\delta_e\simeq\delta_\gamma$ as both electrons and photons behave as relativistic degrees of freedom in thermal equilibrium. The electrons therefore no longer appear as an independent degree of freedom. They only enter the reduced system through the effective rates, i.e. $\Gamma_{\gamma B}+ \Gamma_{\gamma e}$ and their enthalpy $h_e$. The initial conditions read
\begin{equation}
\begin{aligned}
\delta_{\gamma}(\eta_i)&= \delta_{e}(\eta_i) =\delta_{\nu}(\eta_i) \equiv \delta_i,\\
\delta_{B_-}(\eta_i) &\equiv 0,\qquad \delta_{B_+}(\eta_i)=\left(\frac32 + \frac{m_B}{T(\eta_i)}\right)\frac{\delta_{i}}{4},\\
\Phi(\eta_i) &= -\frac{\sum_a\Omega_a\delta_a}{2\left(1+(k\eta_i)^{2}/3\right)},\\
\Phi'(\eta_i) &\equiv 0,\\
v_{\gamma i} &= v_{e i}=v_{\nu i}=v_{B i}
=-\frac{2k\eta_i\Phi(\eta_i)}{3\left(1+\omega\right)}
\;\xrightarrow{\;\omega=1/3\;}\;-\tfrac12 x_i\Phi(\eta_i),\\
S_{B_-}(\eta_i) &= -\tfrac34 \delta_{i}, \qquad S_{B_+}(\eta_i) = \left(-\frac32 + \frac{m_B}{T_i}\right)\frac{\delta_{i}}{4}.
\end{aligned}
\end{equation}
Assuming initially tightly coupled fluids, the velocity relations can be derived from Eq.~\eqref{eq:E0i}, while the initial density conditions are set by Eq.~\eqref{eq:delta_equilibrium}. Here, we assume that there is no initial fluctuation in the asymmetric baryon fluid. In App.~\ref{app:modeldep} we discuss in more detail what happens if this is not the case. $\delta_i$ denotes the initial density contrast and is therefore source dependent. In Fig.~\ref{fig:differentialequations} we evolve the system numerically from the corresponding initial conditions. Around neutrino decoupling the photon and neutrino density fluctuations are washed out, but the baryon isocurvature is preserved. Because baryons and photons remain tightly coupled, the outflow that erases a photon overdensity carries baryons along with it and leaves a compensating baryon underdensity. Neutrino free-streaming erases the metric perturbations only partially, and the surviving potentials act as a late-time source that regenerates the previously damped density and velocity perturbations, as seen in the top-right panel of Fig.~\ref{fig:differentialequations}. We confirm the same behavior of perturbation along with isocurvature conservation in the full system.

\begin{figure}
    \centering
    \includegraphics[width=0.48\linewidth]{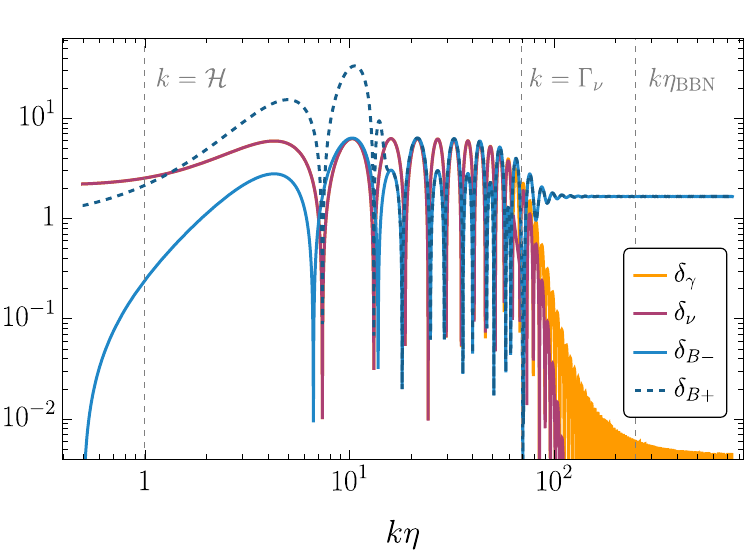}
    \includegraphics[width=0.48\linewidth]{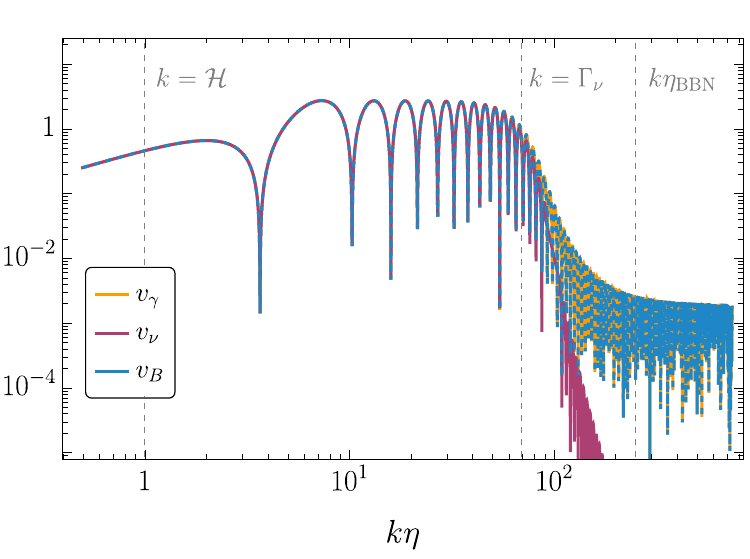}
    \includegraphics[width=0.48\linewidth]{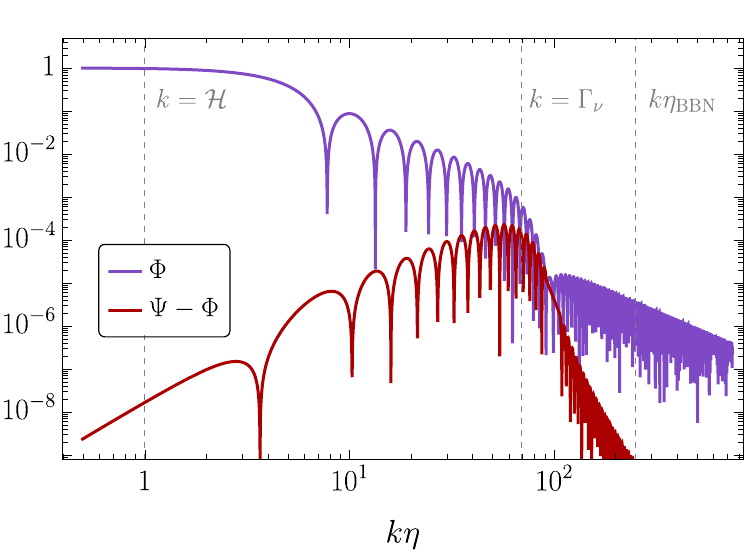}
    \includegraphics[width=0.48\linewidth]{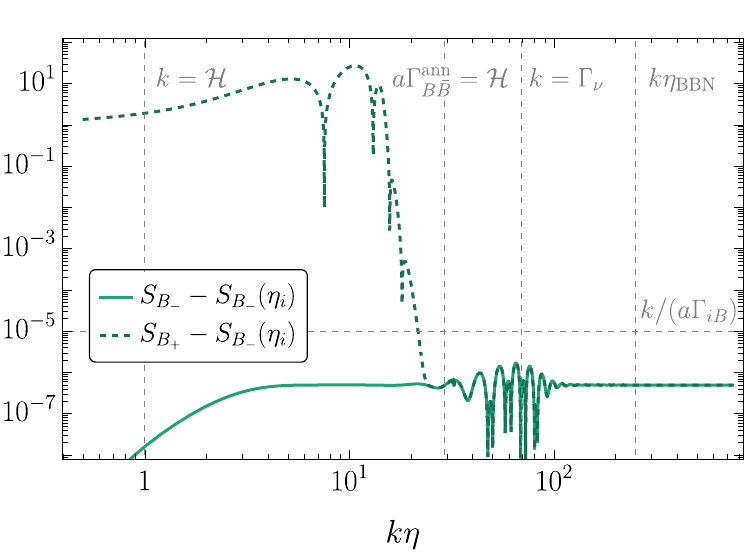}
    \caption{\it Numerical evolution of a single Fourier mode of the coupled $\gamma$--$\nu$--$B$ perturbation system as a function of $k\eta$. \textbf{Top left:} density contrasts of photons ($\delta_\gamma$), neutrinos ($\delta_\nu$), and the net-baryon ($\delta_{B_-}$, solid) and total-baryon ($\delta_{B_+}$, dashed) components. \textbf{Top right:} the corresponding velocities $v_\gamma$, $v_\nu$, $v_B$. \textbf{Bottom left:} the Newtonian potential $\Phi$ and the anisotropic-stress (shear) difference $\Psi-\Phi$. \textbf{Bottom right:} the baryon isocurvature, plotted as the deviation from its conserved initial value $S_{B_-}(\eta_0)$. The net component $S_{B_-}$ (solid) stays conserved, whereas the total component $S_{B_+}$ (dashed) departs during baryon--antibaryon annihilation and relaxes to the same value once annihilation is complete. All quantities are normalized to the primordial potential $\Phi(\eta_i)$. The vertical dashed lines mark, from left to right, horizon crossing $k=\mathcal{H}$ (the transition from super- to sub-horizon evolution), baryon--antibaryon annihilation $a\Gamma_{B\bar B}^{\rm ann}=\mathcal H$, neutrino decoupling and the onset of free-streaming $k\simeq\Gamma_\nu$, and the onset of BBN at $k\eta_{\rm BBN}$. The small residual difference of $S_{B_\pm}$ at late times (the floor in the bottom-right panel) is set by the finite coupling rate, which we fix to $\Gamma/k\lesssim10^5$ (as well as $2a\langle\sigma v\rangle_{B\bar B}^{\rm ann} (n_B^{\rm thermal})^2/(k\bar n_{b+})\lesssim10^5$ with  $\langle\sigma v\rangle_{B\bar B}^{\rm ann}\simeq100~\rm GeV^{-2}$ as an order of magnitude estimate~\cite{Baumann:2022mni}) for numerical stability in the Euler equations. For the much larger physical rates the isocurvature is conserved to even higher accuracy. In this plot, we fixed $g_{\star}\simeq30$ and comoving momentum $k\simeq4.4\times10^{6}~\rm Mpc^{-1} $, which corresponds to at horizon entry at $T\simeq 0.5~\rm GeV$. We evolve the system down to $T\sim1~\rm MeV$. At this stage neutrinos are decoupled from the baryon-photon fluid and should be treated separately.}
    \label{fig:differentialequations}
\end{figure}
\subsection{Interaction Rates}
\noindent
The four interaction rates that dominate the coupling of the plasma around the QCD epoch are: electron--proton Coulomb rate $\Gamma_{eB}$, photon--electron Compton rate $\Gamma_{\gamma e}$, photon--proton Compton rates $\Gamma_{\gamma B}$, and the
neutrino--electron weak rate $\Gamma_{\nu e}$ (see Tab.~\ref{tab:interactions_matrix}). Using the transport
cross section of Ref.~\cite{Suh:1998yi} (valid for $T\lesssim m_p$), the Coulomb rate in the relativistic regime $T\gtrsim m_e$ reads
\begin{align}
\Gamma_{eB}&=a\,\bar n_p\left\langle\sigma_{ep}^{\rm Coul}v_{\rm rel}\right\rangle
=a\,\bar n_p\,\frac{4\pi}{3\zeta(3)}\frac{\alpha_{\rm EM}^2\Lambda(T)}{T^2},
\qquad \bar n_p=\tfrac12\bar n_{B_+},\nonumber\\
\Lambda(T)&\equiv\ln(\lambda_{\rm max}/\lambda_{\rm min}),\qquad
\lambda_{\rm max}\simeq 1/m_D,\qquad \lambda_{\rm min}\simeq 1/T,
\label{eq:Gamma_eb_definition}
\end{align}
where $\Lambda(T)$ is the Coulomb logarithm, set by the range of scales over which the interaction acts, and $m_D=eT/\sqrt3$ is the Debye mass at leading order in thermal QED~\cite{LeBellac:1996}.

Photons scatter off electrons and protons via Compton scattering,
\begin{equation}
\Gamma_{\gamma e}=a\,\bar n_e\,\sigma_{\rm KN}^{e},\qquad
\Gamma_{\gamma b}=a\,\bar n_p\,\sigma_{\rm KN}^{p},\qquad
\bar n_e\equiv\bar n_{e^-}+\bar n_{e^+}=3\zeta(3)T^3/\pi^2,
\end{equation}
with the Klein--Nishina cross section~\cite{rybicki_lightman_1979}, for $i=e,p$,
\begin{align}
\sigma_{\rm KN}^{i}(y)&=\sigma_T^{i}\,\frac34\left[\frac{1+y}{y^3}\!\left(\frac{2y(1+y)}{1+2y}-\ln(1+2y)\right)+\frac{\ln(1+2y)}{2y}-\frac{1+3y}{(1+2y)^2}\right],\nonumber\\
y&=2.7\frac{T}{m_i}\left(3\frac{T}{m_i}+\frac{K_1(m_i/T)}{K_2(m_i/T)}\right),\qquad
\sigma_T^{i}=\frac{8\pi\alpha_{\rm EM}^2}{3m_i^2},
\label{eq:sigma_KN_correct}
\end{align}
where $y$ is the thermally averaged photon energy~\cite{Jeong:2014gna} and $K_\nu$
are modified Bessel functions of the second kind. Neutrinos couple to electrons through the weak rate
\begin{equation}
\Gamma_{\nu e}=C_{\nu e}\,a\,G_F^2 T^5,\qquad C_{\nu e}\sim\mathcal{O}(1),
\end{equation}
valid for $T\ll M_W$, with $G_F$ the Fermi constant.
\begin{table}[ht]
\centering
\renewcommand{\arraystretch}{1.5}
\begin{tabular}{c|cccc}
\hline\hline
$\Gamma_{ab}$ & $\gamma$ & $e$ & $\nu$ & $B$ \\
\hline
$\gamma$ & --      & KN               & --               & KN \\
$e$      & KN      & --               & Weak & Coulomb \\
$\nu$    & --      & Weak & --               & -- \\
$B$      & KN      & Coulomb          & --               & --\\
\hline\hline
\end{tabular}
\caption{\it Interaction rates of the coupled fluid system. ``Coulomb'' is the electron--proton Coulomb rate, ``KN'' the Klein--Nishina (Compton) rate for photons off electrons or protons, and ``weak'' the neutrino--electron rate.}
\label{tab:interactions_matrix}
\end{table}
In tight coupling regime ($\Gamma_a\equiv\sum_b\Gamma_{ab}\gg k$), the anisotropic stress reads $\Pi_a=-(4/9)\,v_a k/\Gamma_a$ ($4/9\to8/15$ for photons to include polarization corrections). This gives
\begin{equation}
\Pi_\gamma=-\frac{8}{15}\frac{k\,v_\gamma}{\Gamma_{\gamma e}+\Gamma_{\gamma B}},
\qquad
\Pi_\nu=-\frac49\frac{k\,v_\nu}{\Gamma_{\nu e}},
\qquad
\Pi_e=-\frac49\frac{k\,v_e}{\Gamma_{e\gamma}+\Gamma_{eB}+\Gamma_{e\nu}}.
\label{eq:Pi_gamma_nu}
\end{equation}
Baryons are non-relativistic at the temperatures of interest, so we set $\Pi_B=0$. Since we work in conformal time, we express all relevant interaction rates in terms of $\eta$ rather than temperature. Assuming radiation domination with, for simplicity, approximately constant relativistic degrees of freedom across the relevant range for the interaction rates, the interaction rates and enthalpies can be rewritten with $T(\eta)\propto1/\eta$. A change in this effective treatment changes the characteristic scales slightly. We verified that this does not change any of our conclusions, and especially baryon isocurvature conservation.

\subsection{Analytic Estimate}
\noindent
As shown in equation Eq.~\eqref{eq:S_B_evolution}, the baryon isocurvature changes with the photon--baryon velocity difference as 
\begin{equation}
  S_B' =k( v_B - v_\gamma),
  \qquad S_B \equiv \delta_B - \tfrac34\delta_\gamma.
\end{equation}
Since the electrons are tightly locked to the
baryons ($v_e \simeq v_B$), the photons scatter off both electrons and baryons,
and the two relevant Euler equations are
\begin{align}
  v_\gamma' &= \Big(-\tfrac14 k\delta_\gamma + \tfrac23 k\Pi_\gamma - k\Psi\Big)
             - \big(\Gamma_{\gamma e}+\Gamma_{\gamma B}\big)\,(v_\gamma - v_B),
             \\[4pt]
  v_B' &= -k\Psi
        - \frac{h_B}{h_e+h_B}\,\mathcal{H}\,v_B
        - \frac{h_e}{h_e+h_B}\,\tfrac14 k\delta_\gamma
        + \frac{h_\gamma}{h_e+h_B}\big(\Gamma_{\gamma e}+\Gamma_{\gamma B}\big)
          (v_\gamma - v_B) .
\end{align}
Subtracting the photon equation from the baryon equation, the common
gravitational force $-k\Psi$ cancels, and the velocity difference obeys
\begin{equation}
  (v_B - v_\gamma)'
  = \Big(\tfrac14 k\delta_\gamma - \tfrac23 k\Pi_\gamma
         - \tfrac{h_e}{h_e+h_B}\tfrac14 k\delta_\gamma
         - \tfrac{h_B}{h_e+h_B}\mathcal{H}v_B\Big)
  - \big(\Gamma_{\gamma e}+\Gamma_{\gamma B}\big)
    \Big(1 + \tfrac{h_\gamma}{h_e+h_B}\Big)(v_B - v_\gamma) .
\end{equation}
If the scattering time
$(\Gamma_{\gamma e}+\Gamma_{\gamma B})^{-1}$ is much shorter compared to the cosmological time scales ($\mathcal H,k$), the velocity slip relaxes to the quasi-stationary attractor $(v_B-v_\gamma)'\simeq 0$ and yields
\begin{equation}
  S_B' = k(v_B - v_\gamma)
  \;\sim\;
  \frac{k^2}{\Gamma_{\gamma e}+\Gamma_{\gamma B}}\,\mathcal{O}\big(\delta_\gamma,\,v_B,\,\dots\big)
\end{equation}
The numerator scales with the evolved initial perturbations, while the denominator is the total rate at which photons scatter off electrons and baryons. {The same suppression appears in the nonlinear treatments of Refs.~\cite{Alcock:1990bb,Jedamzik:1993tcf}. }

\clearpage
\section{Nucleon Diffusion}
\label{app:nucleondiff}
\noindent App.~\ref{app:perturbation} establishes that rapid SM interactions drive the plasma toward local thermal equilibrium while preserving the long-wavelength baryon isocurvature perturbation. Once this stage is reached, the remaining process capable of modifying its spatial structure before BBN is microscopic nucleon diffusion. Since diffusion acts preferentially on short wavelengths, its importance depends on the characteristic scale of the source. The relevant transport changes across weak freeze-out. At temperatures above
\begin{equation}
T_f \simeq 0.8~{\rm MeV},
\label{eq:weakfreezeout}
\end{equation}
weak reactions such as
\begin{equation}
n+e^+ \leftrightarrow p+\bar\nu_e, \qquad n+\nu_e \leftrightarrow p+e^-,
\label{eq:weak_np}
\end{equation}
maintain chemical equilibrium between neutrons and protons. In this regime the neutron and proton abundances remain tightly linked, while the charged component is efficiently coupled to the electron--photon plasma. Relative transport of the baryon number is therefore strongly suppressed.

After weak freeze-out, neutrons and protons no longer track each other chemically. Protons remain coupled to the charged plasma, whereas neutrons, being electrically neutral, can diffuse relative to it. Neutron diffusion, therefore, becomes the dominant transport mechanism for baryon inhomogeneities between weak freeze-out and nucleosynthesis~\cite{Bagherian:2025puf}. Detailed studies of inhomogeneous BBN similarly find that neutrons diffuse over substantially larger distances than protons~\cite{Applegate:1987hm,Jedamzik:1993dc}. We define the comoving diffusion lengths of neutrons and protons by
\begin{equation}
d_i^2(t) \equiv \int_{t_f}^{t} \frac{D_i(t')}{a^2(t')}\,dt', \qquad i=n,p,
\label{eq:diffusionlength}
\end{equation}
where $D_i$ is the corresponding physical diffusion coefficient. The precise numerical normalization depends on the convention adopted for the diffusion kernel and is not important for the parametric discussion below. For neutrons, detailed calculations give a characteristic comoving diffusion length before BBN of order
\begin{equation}
d_n \sim 10^{-9}-10^{-8}\ {\rm Mpc}, \qquad k_n^{\rm diff}\equiv d_n^{-1} \sim 10^8-10^9\ {\rm Mpc}^{-1},
\label{eq:neutrondiffscale}
\end{equation}
with an order-one dependence on the thermal history and on the precise definition of the diffusion scale~\cite{Bagherian:2025puf,Applegate:1987hm,Kurki-Suonio:1992knt,Suh:1998nt}.
The proton diffusion length is smaller,
\begin{equation}
d_p < d_n,
\label{eq:dpdn}
\end{equation}
because Coulomb interactions suppress proton transport. This gives three physically distinct regimes for an isocurvature mode with comoving wavenumber $k$:

1) For sufficiently long wavelengths,
\begin{equation}
k d_n \ll 1,
\label{eq:nodiffusion}
\end{equation}
neither neutron nor proton diffusion significantly modifies the initial inhomogeneity. In this limit
\begin{equation}
S_B(k,t_{\rm BBN}) \simeq S_B(k,t_f),
\label{eq:SBnodiff}
\end{equation}
and the independent homogeneous-patch treatment of Appendix~\ref{app:bbn} applies directly. 

2) At shorter wavelengths there is an intermediate regime,
\begin{equation}
k d_n \gtrsim 1, \qquad k d_p \lesssim 1,
\label{eq:intermediatediff}
\end{equation}
in which neutron diffusion can substantially smooth the neutron density while a proton inhomogeneity remains. In this regime the local state can no longer be described by a single fluctuating baryon-to-photon ratio with the standard homogeneous neutron-to-proton relation. Instead, neutron and proton inhomogeneities must in principle be evolved separately~\cite{Bagherian:2025puf}. Importantly, neutron diffusion alone does not erase the BBN imprint of the baryon inhomogeneity. To verify this explicitly, we performed a simple numerical test with \texttt{PRIMAT}~\cite{Pitrou:2018cgg}. We first evolved the homogeneous plasma down to $T=100\,{\rm keV}$, before substantial nuclear processing, and then divided it into two equal-volume regions. We imposed opposite proton fluctuations while keeping the neutron density homogeneous,
\begin{equation}
n_p^\pm = \bar n_p(1\pm s), \qquad n_n^\pm = \bar n_n,
\label{eq:protonfluctuation}
\end{equation}
adjusting the local baryon densities and abundance fractions consistently. The two regions were then evolved independently through nucleosynthesis and combined only after the nuclear network had frozen out. We find that the final deuterium abundance retains a quadratic dependence on $\langle S_B^2\rangle$. Thus, smoothing the neutron distribution does not remove the sensitivity of BBN to the surviving proton inhomogeneity. With the neutrons smoothed out, helium production no longer depletes the proton excess, so the local deuterium is enhanced, and the isocurvature bound in fact becomes stronger in this regime. To stay conservative, we do not include this enhancement in the limits quoted throughout this work. A complete treatment of this regime would require evolving the neutron and proton spatial distributions continuously together with the nuclear network, as discussed in Ref.~\cite{Bagherian:2025puf}. For the purpose of the present work, the important point is that neutron diffusion should not be identified with complete erasure of the baryon isocurvature signal. 

3) Only once proton diffusion also becomes efficient,
\begin{equation}
k d_p \gtrsim 1,
\label{eq:protonwashout}
\end{equation}
can the baryon-number inhomogeneity itself be efficiently washed out before nucleosynthesis. A useful criterion for the survival of a BBN imprint is
therefore
\begin{equation}
k_{\rm source}\,d_p \lesssim 1,
\label{eq:survivalcriterion}
\end{equation}
whereas the stronger condition
\begin{equation}
k_{\rm source}\,d_n \ll 1
\label{eq:simpleBBNcriterion}
\end{equation}
guarantees that the simpler homogeneous-patch calculation of App.~\ref{app:bbn} can be applied without resolving neutron and proton transport separately. 

\clearpage
\section{BBN Limits on Baryon Isocurvature}
\label{app:bbn}
\noindent
We now relate the surviving baryon isocurvature to the primordial deuterium abundance. The calculation rests on two assumptions. First, we restrict to scales on which nucleon diffusion is negligible, Eq.~\eqref{eq:simpleBBNcriterion}, so that each region burns at its own fixed baryon-to-photon ratio. Second, following the evolution described in App.~\ref{app:perturbation}, we neglect residual photon-temperature perturbations during nucleosynthesis, so that the local $\eta_B$ is the only label a region carries into BBN. 

\subsection{Effect of Baryon Inhomogeneities}
\noindent
Baryon isocurvature produces a spatially varying baryon-to-photon ratio,
\begin{equation}
\eta_B(\mathbf{x})=\bar{\eta}_B[1+S_B(\mathbf{x})],
\qquad \langle S_B\rangle=0,
\label{eq:eta_of_x}
\end{equation}
where angular brackets denote a volume average. Each region therefore burns to a different deuterium abundance. Since deuterium and hydrogen are measured in absorbers that sample the mixed material, the observable is the ratio of the averaged number densities rather than the average of the local ratio~{\cite{Bagherian:2025puf}},
\begin{equation}
Y_D^{\rm S}
=\frac{\langle n_D\rangle}{\langle n_H\rangle}
=\frac{\langle n_HY_D(\eta_B)\rangle}{\langle n_H\rangle},
\label{eq:deuterium_mixing}
\end{equation}
where $Y_D(\eta_B)=n_D/n_H$ is the local abundance and the superscript S labels the value in the presence of isocurvature. Hydrogen-rich regions therefore carry more weight. At approximately uniform photon temperature, and neglecting the weak dependence of the hydrogen fraction on $\eta_B$, that weight is $n_H\propto\eta_B$, so by Eq.~\eqref{eq:eta_of_x} the denominator of Eq.~\eqref{eq:deuterium_mixing} is exactly $\bar n_H$. Expanding the numerator to second order in $S_B$ gives
\begin{equation}
\frac{Y_D^{\rm S}-Y_D^0}{Y_D^0}
=C_D\langle S_B^2\rangle+\mathcal{O}(S_B^3),
\qquad
Y_D^0\equiv Y_D(\bar{\eta}_B),
\label{eq:deuterium_response}
\end{equation}
with
\begin{equation}
C_D\equiv
\frac{1}{2Y_D^0}
\left[
\frac{d^2Y_D}{d(\ln\eta_B)^2}
+\frac{dY_D}{d\ln\eta_B}
\right]_{\bar{\eta}_B}.
\label{eq:CD_general}
\end{equation}
The term linear in $S_B$ averages to zero, so the leading effect is set by the variance, and it survives only because $Y_D$ depends nonlinearly on $\eta_B$. Near the cosmological baryon density, the local abundance is well described by a power law,
\begin{equation}
Y_D(\eta_B)\simeq
Y_D^0\left(\frac{\eta_B}{\bar{\eta}_B}\right)^{\alpha_D}.
\label{eq:YD_powerlaw}
\end{equation}
Our \texttt{PRIMAT} calculation gives $\alpha_D=-1.64$ at $\bar\eta_B=6.10\times10^{-10}$. Within this power-law approximation, Eq.~\eqref{eq:CD_general} reduces to
\begin{equation}
C_D=\frac{\alpha_D(\alpha_D+1)}{2}
\simeq0.525.
\label{eq:CD_powerlaw}
\end{equation}
Since $C_D>0$, baryon inhomogeneities raise the mean deuterium abundance at leading order. {The sign is not an artefact of the power-law fit but follows from Jensen's inequality applied to the numerator of Eq.~\eqref{eq:deuterium_mixing}~\cite{Barrow:2018yyg}. The same argument applied to $^7$Li shows that inhomogeneities worsen rather than relieve the lithium problem~\cite{Alcock:1987tx,Kurki-Suonio:1996wfr,Jedamzik:2001qc}.}
More generally, defining $\alpha=d\ln Y_D/d\ln\eta_B$, Eq.~\eqref{eq:CD_general} gives
\begin{equation}
C_D=\frac12
\left[
\alpha(\alpha+1)+\frac{d\alpha}{d\ln\eta_B}
\right]_{\bar{\eta}_B}.
\end{equation}
{Keeping the running of $\alpha_D$ adds $d\alpha_D/d\ln\eta_B\simeq-0.32$ to the bracket of Eq.~\eqref{eq:CD_general} and lowers $C_D$ to $0.36$. Higher orders in $S_B$ push the exact $\langle D\rangle/\langle H\rangle$ back up, so both values track it to within the spread between nuclear prescriptions. We therefore keep the closed form $C_D=0.525$, whose accuracy is quoted below.}

\subsection{Observational Abundance and BBN Predictions}
\noindent
Measurements of primordial deuterium give somewhat different central values and uncertainties. We adopt the value recommended by the 2026 Particle Data Group review~\cite{ParticleDataGroup:2026mpi},
\begin{equation}
Y_D^{\rm PDG}=(2.508\pm0.029)\times10^{-5}.
\label{eq:deuterium_PDG}
\end{equation}
This value combines twelve measurements, among them Refs.~{\cite{Cooke:2017cwo,Kislitsyn:2024jvk}}, with the uncertainty enlarged slightly to account for their scatter.

BBN predictions depend in addition on the treatment of the nuclear reaction rates, in particular on which data are included and how the cross sections are fitted. The four prescriptions we compare differ in exactly that choice:
\begin{itemize}
    \item \texttt{PRIMAT}\textbf{:} theory-based cross-section shapes normalized to experimental data~\cite{Pitrou:2018cgg}. It predicts a lower $Y_D$ than observed.
    \item \texttt{PArthENoPE}\textbf{:} low-degree polynomial fits to the cross-section data~\cite{Gariazzo:2021iiu}. It predicts a $Y_D$ close to the observed abundance.
    \item \texttt{PRyMordial}\textbf{:} the nuclear inputs of Ref.~\cite{Yeh:2020mgl}, which incorporate the post-LUNA determination of the $d(p,\gamma)^3$He rate~\cite{Burns:2023sgx}.
    \item \textbf{Gaussian processes:} a non-parametric fit to the nuclear cross sections~\cite{Launders:2026ciu}.
\end{itemize} 
The spread in the predicted $D/H$ therefore measures the uncertainty in the deuterium-burning rates rather than any cosmological uncertainty. The four predictions are collected in Table~\ref{tab:bbn_predictions}, with the quoted uncertainties from Ref.~\cite{Launders:2026ciu}. As in the main text, we use \texttt{PRIMAT} and \texttt{PArthENoPE} for our main comparison, and show the other two for reference.
\begin{table}[t]
\centering
\begin{tabular}{lc}
\hline\hline
Nuclear prescription & $10^5Y_D^0$\\
\hline
\texttt{PRIMAT}            & $2.444\pm0.037$\\
\texttt{PArthENoPE}       & $2.512\pm0.042$\\
\texttt{PRyMordial}        & $2.520\pm0.104$\\
Gaussian processes & $2.442\pm0.040$\\
\hline\hline
\end{tabular}
\caption{\it Homogeneous predictions from Table~II of Ref.~\cite{Launders:2026ciu}, evaluated with a common BBN solver and cosmological inputs. Errors are one standard deviation.}
\label{tab:bbn_predictions}
\end{table}

\subsection{Comparison with Baryon Isocurvature}
\noindent
We adopt the \texttt{PRIMAT}-derived coefficient $C_D=0.525$ for all four prescriptions,
\begin{equation}
Y_{D,i}^{\rm S}(S_B)
\simeq Y_{D,i}^0(1+C_D\langle S_B^2\rangle),
\label{eq:deuterium_prediction_isocurvature}
\end{equation}
rather than repeating the inhomogeneous calculation for each of them. We verify below that using each network's own slope in Eq.~\eqref{eq:CD_powerlaw} moves the results by less than $1\%$. Starting from each central prediction $Y_{D,i}^0$ of Table~\ref{tab:bbn_predictions}, we then ask which amplitudes bring that prediction into agreement with the PDG value of Eq.~\eqref{eq:deuterium_PDG} at the $N\sigma$ level,
\begin{equation}
\left|
Y_{D,i}^0(1+C_D\langle S_B^2\rangle)-Y_D^{\rm PDG}
\right|\leq N\,\sigma_{{\rm comb},i}.
\label{eq:deuterium_agreement}
\end{equation}
The tolerance combines the observational error of Eq.~\eqref{eq:deuterium_PDG} with the prediction error of Table~\ref{tab:bbn_predictions},
\begin{equation}
\sigma_{{\rm comb},i}
\equiv\sqrt{\sigma_{\rm obs}^2+\sigma_{{\rm pred},i}^2},
\qquad
\sigma_{\rm obs}=0.029\times10^{-5},
\label{eq:sigma_comb}
\end{equation}
which we take to be independent and Gaussian, and we hold $\sigma_{{\rm pred},i}$ fixed at its homogeneous value. Writing $\Delta_i=Y_D^{\rm PDG}-Y_{D,i}^0$, Eq.~\eqref{eq:deuterium_agreement} is solved by
\begin{equation}
\max\left[0,\frac{\Delta_i-N\sigma_{{\rm comb},i}}{C_DY_{D,i}^0}\right]
\leq \langle S_B^2\rangle
\leq
\frac{\Delta_i+N\sigma_{{\rm comb},i}}{C_DY_{D,i}^0},
\label{eq:isocurvature_interval}
\end{equation}
whenever the upper endpoint is nonnegative, and has no solution otherwise. Table~\ref{tab:bbn_isocurvature} lists the square roots of these endpoints for $N=1$ and $N=2$, and Fig.~\ref{fig:two_codes_crosscheck} shows the same construction graphically.
\begin{table}[t]
\centering
\begin{tabular}{lcc}
\hline\hline
Nuclear prescription &
 $1\sigma$&
 $2\sigma$\\
\hline
\texttt{PRIMAT}             &  ${0.12}$--$0.29$ & $0$--$0.35$ \\
\texttt{PArthENoPE}         &  $0$--${0.19}$ & $0$--${0.27}$\\
\texttt{PRyMordial}     &  $0$--${0.27}$ &  $0$--${0.39}$\\
Gaussian processes &  ${0.11}$--$0.30$ &  $0$--${0.36}$\\
\hline\hline
\end{tabular}
\caption{\it Ranges of $\sqrt{\langle S_B^2\rangle}$ that bring each central BBN prediction into agreement with the PDG deuterium abundance of Eq.~\eqref{eq:deuterium_PDG}, from Eq.~\eqref{eq:isocurvature_interval} with $N=1$ and $N=2$. The tolerance is the combined error of Eq.~\eqref{eq:sigma_comb}, so the two columns are not the observational band alone. All central values and errors are those of Ref.~\cite{Launders:2026ciu}, collected in Table~\ref{tab:bbn_predictions}.}
\label{tab:bbn_isocurvature}
\end{table}

The \texttt{PRIMAT} and Gaussian-process central predictions lie below the band, so a positive isocurvature contribution first improves the agreement and only later spoils it. Their intervals accordingly open at a nonzero amplitude, the lower endpoint marking where the shifted prediction enters the band. At $N=2$ the band is already reached at zero amplitude, which is why those lower endpoints vanish. The \texttt{PArthENoPE} and \texttt{PRyMordial} central predictions start inside the band, so for them the isocurvature is bounded from above only.

{Running the \texttt{PRIMAT}, \texttt{PRyMordial} and \texttt{PArthENoPE} networks ourselves reproduces the central values of Table~\ref{tab:bbn_predictions} to two significant figures, and replacing $C_D$ by each network's own slope moves the endpoints by less than $1\%$. The leading-order truncation of Eq.~\eqref{eq:deuterium_prediction_isocurvature} is accurate to $0.6\%$ for $\sqrt{\langle S_B^2\rangle}\lesssim0.3$, below the spread between nuclear prescriptions.}

{Ref.~\cite{Inomata:2018htm} sets a similar bound, but averages $D/H$ rather than the observable $\langle D\rangle/\langle H\rangle$ of Eq.~\eqref{eq:deuterium_mixing}, so their response coefficient differs from ours. In our convention their limit reads $\sqrt{\langle S_B^2\rangle}<0.25$ at $2\sigma$, close to our \texttt{PArthENoPE} limit in Table~\ref{tab:bbn_isocurvature}. Ref.~\cite{Bagherian:2025puf} quotes $\sqrt{\langle S_B^2\rangle}\lesssim0.28$ at $1\sigma$ with \texttt{PRyMordial}, and our \texttt{PRyMordial} row gives $0.27$, which checks our pipeline.}

\begin{figure}[t!]
\centering
\includegraphics[width=\linewidth]{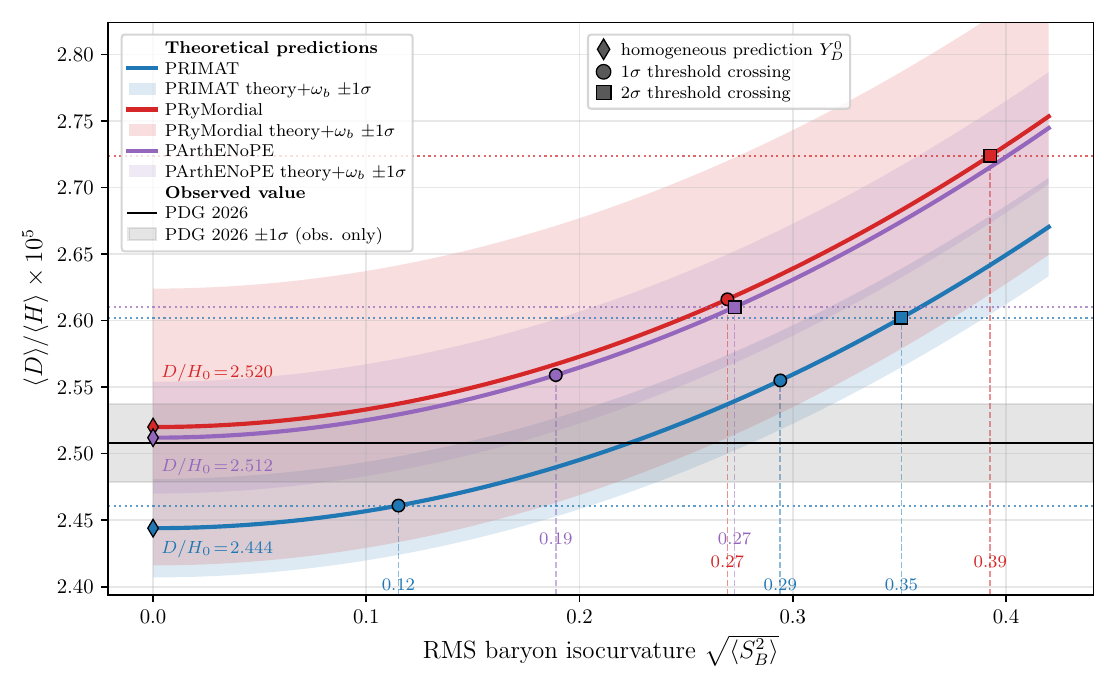}
\caption{\it Mean deuterium abundance $\langle D\rangle/\langle H\rangle$ as a function of the isocurvature amplitude $\sqrt{\langle S_B^2\rangle}$, for three of the four nuclear prescriptions collected in Table~\ref{tab:bbn_predictions}: \texttt{PRIMAT} (blue), \texttt{PRyMordial} (red) and \texttt{PArthENoPE} (purple). The Gaussian-process prescription is omitted because its central value lies within $0.1\%$ of \texttt{PRIMAT} and its curve would be indistinguishable. Each curve starts from that prescription's homogeneous prediction $Y_D^0$ (diamond at $\sqrt{\langle S_B^2\rangle}=0$) and rises quadratically following Eq.~\eqref{eq:deuterium_prediction_isocurvature}, and the shaded band around it is that prescription's own $1\sigma$ prediction error, combining nuclear rates{, the neutron lifetime} and the Planck baryon density $\Omega_B$. The black line and grey band are the PDG abundance and its observational $1\sigma$ interval~\cite{ParticleDataGroup:2026mpi}, Eq.~\eqref{eq:deuterium_PDG}. Dotted lines mark each prescription's combined thresholds $N\sigma_{{\rm comb},i}$ of Eq.~\eqref{eq:sigma_comb} for $N=1,2$, with circles and squares where its curve crosses them. \texttt{PRIMAT} starts below the observational band, so its agreement interval opens at a nonzero $\sqrt{\langle S_B^2\rangle}$, whereas \texttt{PRyMordial} and \texttt{PArthENoPE} start inside it.}
\label{fig:two_codes_crosscheck}
\end{figure}

\clearpage
\section{PTA analysis}
\label{app:pta}
\noindent
We obtain the PTA posteriors with \texttt{PTArcade}~\cite{Mitridate:2023oar}
(v1.1.2) run in \texttt{ceffyl} mode~\cite{Lamb:2023jls} without accounting for spatial pulsar correlations and use the NANOGrav 15-year data~\cite{NANOGrav:2023gor}. We do not include a supermassive black-hole binary component and assume that a single new-physics source accounts for the entire common-spectrum process. The chains are generated with \texttt{PTMCMCSampler}~\cite{justin_ellis_2017_1037579} for
$5\times10^{6}$ requested steps. The two-dimensional posteriors are marginalized over the remaining model parameters and displayed at $68\%$ and $95\%$ confidence levels.

Separate analyses are performed for DW annihilation and FOPTs.
Below, we summarize the GW spectra templates used in this work along with their priors and median posterior values for the respective model parameters.

\subsection{Domain Walls}
\noindent
We adopt several GW templates for the annihilating DW network, denoted NG~\cite{NANOGrav:2023hvm,Saikawa:2017hiv}, NRT~\cite{Notari:2025kqq}, DGRV~\cite{Dankovsky:2026kjj} and CCB~\cite{Cyr:2025nzf}. The GW spectrum can be parameterized as
\begin{equation}
\label{eq:GW_DW}
    \Omega^{\rm ann}_{\rm GW}(k)\simeq\frac{3}{32\pi}\,\epsilon\,\alpha_{\rm DW}(\eta_{\rm ann})^2\times \mathcal S\left(\frac{k}{2\pi\mathcal H(\eta_{\rm ann})}\right), \qquad {\rm with} \qquad \mathcal S(x)=\frac{(a+b)^c}{\left(b\left(\frac{x}{x_p}\right)^{-\frac{a}{c}}+a\left(\frac{x}{x_p}\right)^{\frac{b}{c}}\right)^c},
\end{equation}
where $\mathcal H=aH$, $\eta_{\rm ann}$ denotes the epoch at which GW production ceases, and $\mathcal S$ is normalized to unity at the peak~\footnote{The parameterization of Ref.~\cite{Hiramatsu:2013qaa} is instead normalized at $x=1$ but peaks at $x=3^{1/4}$, which shifts the inferred $T_{\rm ann}$ by $\sim30\%$ at the same $(\alpha,\beta,\delta)$.}. The templates of NRT~\cite{Notari:2025kqq}, DGRV~\cite{Dankovsky:2026kjj} and CCB~\cite{Cyr:2025nzf} all agree with Eq.~\eqref{eq:GW_DW}, and we compare their best fit values for a biased $\phi^4$ potential with the values used by the NANOGrav collaboration
\begin{table}[h]
\centering
\begin{tabular}{lcccc}
\hline\hline
 & $a$ & $b$ & $c$ & $\epsilon$ \\
\hline
NG $-$ $b,c$ fixed & $3$ & $1$ & $1$ & $0.7$ \\
NG $-$ $b,c$ free  & $3$ & free & free & $0.7$ \\
NRT   & $3$ & $0.80\pm0.08$ & $1.52\pm0.46$ & $0.078\pm0.007$ \\
DGRV  & $2.6\pm0.2$ & $0.85\pm0.01$ & $0.90\pm0.10$ & $0.17$ \\
CCB   & $3$ & $1.56\pm0.20$ & $1$ & free \\
\hline\hline
\end{tabular}
\caption{\it Spectral-shape parameters of the DW GW templates entering Eq.~\eqref{eq:GW_DW}.}
\label{tab:DWtemplates}
\end{table}
Here $a=3$ is fixed by causality, except in DGRV~\cite{Dankovsky:2026kjj}, which leave it free and find a shallower near-peak IR slope. Since only the combination $x/x_p=k/k_{\rm peak}$ enters $\mathcal S$, the value of $x_p$ does not affect the spectral shape but fixes the map $f_p=x_p f_H(T_{\rm ann})$. NRT~\cite{Notari:2025kqq} find $x_p\simeq2.32$ for the annihilating network, in contrast to the value $x_p=1$ commonly assumed. Both NRT~\cite{Notari:2025kqq} and CCB~\cite{Cyr:2025nzf} further find that emission continues past the commonly assumed annihilation timescale of Eq.~\eqref{eq:sigmaDW_App}. This yields a higher relative energy fraction at annihilation and an enhanced GW signal.

With this and imposing uniform (log) priors on the model parameters
\begin{equation}
\log_{10}(\sigma^{1/3}/{\rm GeV})\in[3,6],\qquad
\log_{10}(T_{\rm ann}/{\rm GeV})\in[-4,0], \qquad b\in[0.5,1], \qquad c\in[0.3,3],
\end{equation}
where the slope priors on $b$ and $c$ apply only to the NG $-$ $b,c$ free run, every other template having its shape parameters fixed to the values tabulated above. We summarize the median posterior values, with the $16$th and $84$th percentiles of the marginalized posterior in Tab.~\ref{tab:medians_DW} and show the $1\sigma$ posteriors in Fig.~\ref{fig:DWApp}.
\begin{table}[h!]
\begin{center}
\renewcommand{\arraystretch}{1.4}
\begin{tabular}{lcc}
\hline\hline
& \ \ $T_{\rm ann}$ [GeV] \ \ & \ \ $\alpha_{\rm DW}$ \ \ \\
\hline
NG $-$ $b,c$ fixed & $0.128^{+0.030}_{-0.027}$ & $0.147^{+0.063}_{-0.043}$ \\
NG $-$ $b,c$ free  & $0.185^{+0.108}_{-0.060}$ & $0.160^{+0.088}_{-0.055}$ \\
NRT                         & $0.116^{+0.042}_{-0.031}$ & $0.133^{+0.041}_{-0.036}$ \\
DGRV                        & $0.205^{+0.075}_{-0.044}$ & $0.104^{+0.056}_{-0.035}$ \\
CCB                         & $0.182^{+0.035}_{-0.031}$ & $0.178^{+0.076}_{-0.041}$ \\
\hline\hline
\end{tabular}
\caption{\it Median values of the posterior distributions of the DW parameters.}
\label{tab:medians_DW}
\end{center}
\end{table}
The best fit values for the slopes in the corresponding NG template are $b \simeq 0.741_{-0.166}^{+0.173}$ and $c \simeq 2.069_{-0.926}^{+0.653}$. Neither slope is informative, since the $b$ posterior is flat across its prior $[0.5,1]$ and $c$ only mildly prefers the upper end of $[0.3,3]$. For every template $\alpha_{\rm DW}$ is quoted with the common scaling area parameter $\mathcal A=0.8$, so that the posteriors differ by their GW spectrum alone. Adopting instead the wall energy density $\rho_{\rm wall}=2\xi\gamma\sigma H$ measured by DGRV~\cite{Dankovsky:2026kjj}, with $\xi\simeq1.2$, would raise their $\alpha_{\rm DW}$ by $\xi\gamma/\mathcal A\simeq1.6$ at fixed $\sigma$.

\begin{figure}[t!]
    \centering
    \includegraphics[width=0.49\linewidth]{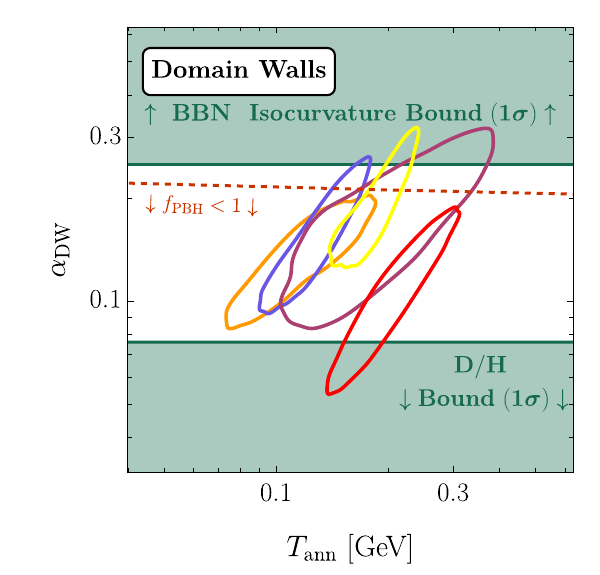}
    \includegraphics[width=0.49\linewidth]{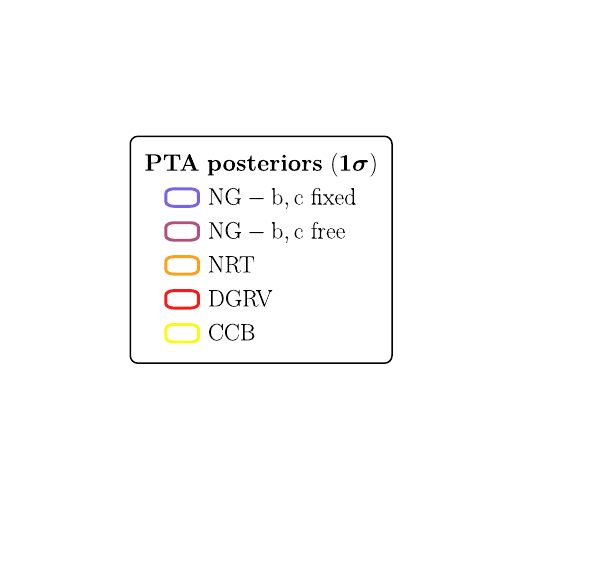}
    \caption{\it BBN baryon-isocurvature exclusion (green shaded, excluded at $1\sigma$) for domain walls, compared with the NANOGrav 15-year posteriors ($1\sigma$) obtained with the GW templates Saikawa~\cite{Hiramatsu:2013qaa,Gouttenoire:2023ftk}, NG with $b,c$ fixed and with $b,c$ free~\cite{NANOGrav:2023hvm}, NRT~\cite{Notari:2025kqq}, DGRV~\cite{Dankovsky:2026kjj} and CCB~\cite{Cyr:2025nzf}, whose parameters are collected in Tab.~\ref{tab:DWtemplates}. The isocurvature bound is a priori  independent of the template, so the spread between the posteriors reflects the current theoretical uncertainty on the DW spectrum alone.}
    \label{fig:DWApp}
\end{figure}

\subsection{First-Order Phase Transitions}
\noindent
A cosmological first-order phase transition sources GWs through three mechanisms: collisions of the vacuum bubble walls, sound waves in the plasma, and magnetohydrodynamic turbulence~\cite{Caprini:2015zlo,Hindmarsh:2015qta}. We adopt the templates of~\cite{Lewicki:2025hxg} and the LISA Cosmology Working Group~\cite{Caprini:2024hue}. We denote the Hubble parameter evaluated at nucleation with $H_n$ and at percolation with $H_p$. For strong transitions, these two differ and the GW signal is dominated by bubble collisions or highly relativistic fluid shells. The spectral shape is well described with the same broken power law as in Eq.~\eqref{eq:GW_DW}~\cite{Lewicki:2025hxg},
\begin{equation}
\label{eq:GW_FOPT_BPL}
\Omega_{\rm GW,\star}^{\rm BPL}(f)= A\left(\frac{H_n}{\beta}\right)^2 \,\mathcal S\!\left(\frac{f}{f_p}\right),
\end{equation}
where we adopt the bulk flow model~\cite{Lewicki:2025hxg} with $(a,b,c)\simeq (1,2,1)$. Accounting for the cosmic expansion, the peak amplitude and frequency scale with $\beta$ as
\begin{equation}
A\simeq0.06\left[1+0.8\left(1-e^{\sqrt{H_n/\beta}}\right)\right],\qquad
\frac{2\pi f_p}{\mathcal H_p}\simeq0.7\frac{\beta}{H_n}\left[1+1.8\left(\frac{\beta}{H_n}\right)^{-1.2}\right],
\end{equation}
evaluated at the percolation time, and reducing to the standard result for $\beta/H_n\gg1$.

Sound waves and turbulence instead dominate for weaker FOPT and $H_\star \equiv H_n\simeq H_p$ follow the double broken power law 
\begin{equation}
\label{eq:GW_FOPT_DBPL}
\Omega_{\rm GW,\star}^{\rm DBPL}(f)=\Omega_{\rm int}\left(\frac{f}{f_1}\right)^{n_1}
\left[1+\left(\frac{f}{f_1}\right)^{a_1}\right]^{\frac{n_2-n_1}{a_1}}
\left[1+\left(\frac{f}{f_2}\right)^{a_2}\right]^{\frac{n_3-n_2}{a_2}},
\end{equation}
with the causal low-frequency slope $n_1=3$ and the remaining slopes summarized below. The peak amplitude $\Omega_{\rm int}$ and characteristic frequencies $f_{1,2}$ are fixed by the thermodynamic parameters $(\alpha,\beta/H_\star,T_{\rm reh},\xi_{\rm W},\epsilon_{\rm turb})$ and differ for sound waves and turbulence. Here, $\xi_{\rm W}$ and $\epsilon_{\rm turb}$ describe the sound wall thickness and the energy fraction in MHD turbulence. We use the  expressions of Ref.~\cite{Caprini:2024hue} throughout, to which we refer for details, each multiplied by the redshift factor $h^2F_{\rm GW,0}$ of that reference. The best fit spectral shape parameters for the respective sources read~\cite{Caprini:2024hue}
\begin{table}[h]
\begin{center}
\renewcommand{\arraystretch}{1.3}
\begin{tabular}{llccccc}
\hline\hline
GW source & template & $n_1$ & $n_2$ & $n_3$ & $a_1$ & $a_2$ \\
\hline
sound waves & DBPL & $3$   & $1$    & $-3$   & $2$   & $4$ \\
turbulence  & DBPL & $3$   & $1$    & $-8/3$ & $4$   & $2.15$ \\
\hline\hline
\end{tabular}
\end{center}
\caption{\it Spectral-shape parameters of the double broken power law, Eq.~\eqref{eq:GW_FOPT_DBPL}. The bulk flow template of Eq.~\eqref{eq:GW_FOPT_BPL} uses $(a,b,c)\simeq(1,2,1)$.}
\end{table}
Imposing uniform log priors,
\begin{equation}
\log_{10}(\beta/H_\star)\in[0,3],\qquad
\log_{10}(T_{\rm reh}/{\rm GeV})\in[-4,1],\qquad
\log_{10}\alpha\in[-2,3],
\end{equation}
for the two LISA templates. For the bulk flow we instead restrict $\beta/H_n\in[4,64]$, the range covered by the simulations behind the suppression fit of Ref.~\cite{Lewicki:2025hxg}, outside of which the fitted amplitude is not defined. We find the following median posterior values, with the $16$th and $84$th percentiles of the marginalized posterior as errors, for the three GW sources in Tab.~\ref{tab:MediansFOPT} and show the $1\sigma$ posteriors in Fig.~\ref{fig:FOPTApp}.
\begin{table}[h]
\begin{center}
\renewcommand{\arraystretch}{1.4}
\begin{tabular}{lccc}
\hline\hline
& \ \ bulk flow \ \ & \ \ sound waves \ \ & \ \ turbulence \ \ \\
\hline
$\beta/H_\star$         & $8.33^{+3.67}_{-3.06}$    & $18.0^{+15.2}_{-13.3}$    & $4.93^{+3.36}_{-2.80}$ \\
$T_{\rm reh}$ [GeV]      & $0.232^{+0.131}_{-0.071}$ & $0.045^{+0.142}_{-0.028}$ & $0.106^{+0.195}_{-0.061}$ \\
$\alpha$            & $32.3^{+305}_{-29.5}$     & $13.9^{+239}_{-13.0}$     & $26.1^{+285}_{-23.7}$ \\
\hline\hline
\end{tabular}
\end{center}
\caption{\it Median values of the posterior distributions of the FOPT parameters.}
\label{tab:MediansFOPT}
\end{table}

It should be noted here that in realistic models of very strong FOPT, such as supercooled phase transitions, there is typically a correlation between $\alpha$ and $\beta/H_\star$~\cite{Goncalves:2025uwh,Balan:2025uke,Kierkla:2025qyz,Kierkla:2025vwp,Bringmann:2026xcx,Biondini:2026uds,Puchades-Ibanez:2026ksh}. This can impact the preferred region of model parameter space and the corresponding values of $\beta/H_\star$ in the fit.  It would therefore be interesting to perform the analysis in concrete models of strong FOPT. 

\begin{figure}[t]
    \centering
    \includegraphics[width=0.49\linewidth]{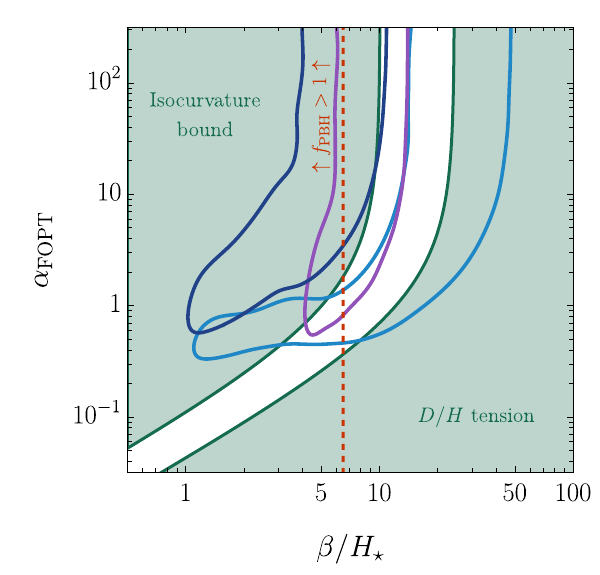}
    \includegraphics[width=0.49\linewidth]{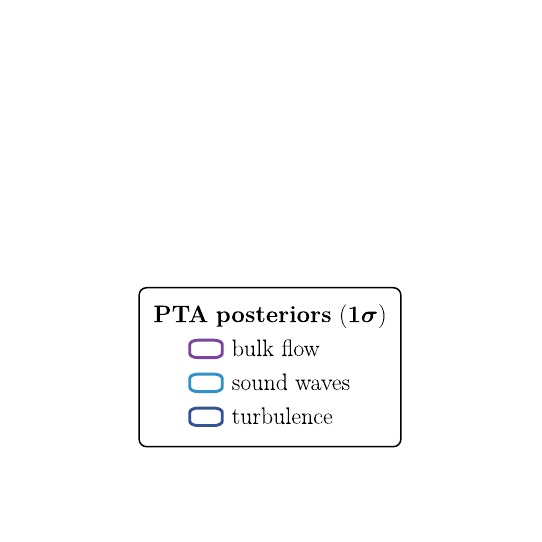}
    \includegraphics[width=0.49\linewidth]{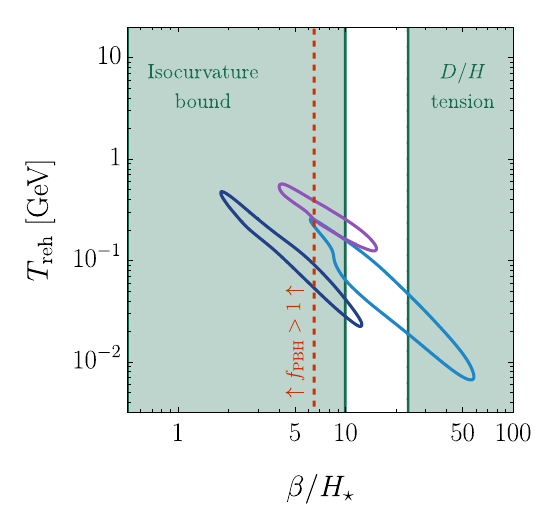}
    \includegraphics[width=0.49\linewidth]{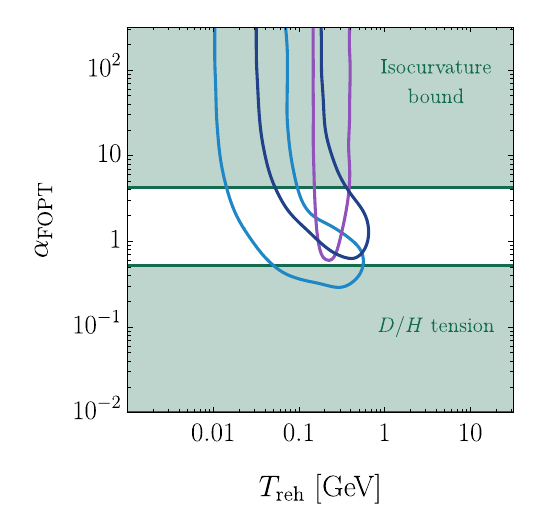}
    \caption{\it  BBN baryon-isocurvature exclusion (green shaded, excluded), computed with \texttt{PRIMAT}, for a first-order phase transition, shown in all three pairwise projections of the model parameter space $(\beta/H_\star,\,\alpha_{\rm FOPT},\,T_{\rm reh})$ and compared with the NANOGrav 15-year posteriors ($1\sigma$). \textbf{Top left:} the $(\beta/H_\star,\alpha_{\rm FOPT})$ plane, where the bound $\beta/H_\star\gtrsim10.3\,\alpha/(1+\alpha)$ acts directly. \textbf{Bottom left:} the $(\beta/H_\star,T_{\rm reh})$ plane. \textbf{Bottom right:} the $(T_{\rm reh},\alpha_{\rm FOPT})$ plane. Because the isocurvature bound is independent of the reheating temperature, in each projection the remaining parameter is fixed to its posterior-averaged value. The GW templates are the strongly supercooled bulk-flow spectrum of Ref.~\cite{Lewicki:2025hxg}, and, for sound waves and turbulence, the LISA Cosmology Working Group parameterizations~\cite{Caprini:2024hue}. The latter are built on the lattice simulations and sound shell model of Refs.~\cite{Hindmarsh:2013xza,Hindmarsh:2015qta,Hindmarsh:2017gnf,Hindmarsh:2016lnk,Hindmarsh:2019phv} for sound waves, and on the MHD turbulence calculations of Refs.~\cite{RoperPol:2022iel,RoperPol:2023bqa} for turbulence.}
    \label{fig:FOPTApp}
\end{figure}

\clearpage
\section{Generalizations Beyond Visible Energy Injection}
\label{app:modeldep}
\noindent
The analysis so far assumes that DW and FOPT do not produce net baryon number and reheat to SM particles. Here, we briefly discuss how our framework is modified if these assumptions are relaxed.

\subsection{Baryon-number Violating Energy Injection}
\noindent
The discussion in the main text assumes that the energy injection does not generate a net baryon asymmetry. One can, however, consider cases where the energy injection itself violates baryon number, for example in late baryogenesis scenarios (i.e., particle production across the bubble wall in the case of FOPTs). In these cases, the local baryon asymmetry $n_B-n_{\bar B}$ becomes spatially inhomogeneous.

Assuming the mechanism generates both entropy and baryon number, the resulting isocurvature perturbations can be parameterized as
\begin{equation}
\delta_B=\xi_B\,\delta_X, \qquad \delta_\gamma=\xi_\gamma\,\delta_X,
\end{equation}
where $\delta_X$ denotes the spatial fluctuation of the source and the coefficients $\xi_B$ and $\xi_\gamma$ encode the microscopic baryogenesis mechanism. The initial baryon isocurvature perturbation therefore becomes
\begin{equation}
S_B = \delta_B-\frac34\delta_\gamma = \left( \xi_B-\frac34\xi_\gamma \right)\delta_X.
\end{equation}
The subsequent evolution is identical to that discussed in App.~\ref{app:perturbation}. Once the energy has thermalized, the plasma no longer retains information about how the perturbation was generated. The microscopic baryogenesis mechanism affects only the initial matching condition for $S_B$, while the constraints of hydrodynamic evolution, nucleon diffusion, and BBN remain unchanged.

\subsection{Dark Radiation and $\Delta N_{\rm eff}$}
\label{app:DR}
\noindent
If the BSM source decays dominantly into a dark sector ($f_{\rm vis} \to 0$), the photon bath remains unperturbed ($\delta_\gamma \to 0$) and $S_B \to 0$. Consequently, the BBN isocurvature bound vanishes for $f_{\rm vis} \ll 1$. In contrast to the previous examples, the injected energy never thermalizes with the visible plasma. The baryon-to-photon ratio therefore remains unchanged, and the dominant cosmological signature is no longer baryon isocurvature but an increase in the total radiation density.

However, stable relativistic dark particles produced in the decay persist as dark radiation, contributing to the effective number of neutrino species $\Delta N_{\rm eff}$. The energy density of dark radiation is:
\begin{equation}
\rho_{\rm dark} =(1 - f_{\rm vis}) \alpha_S  \rho_{\rm rad}.
\end{equation}
Redshifting down to the epoch of photon decoupling yields:
\begin{equation}
\Delta N_{\rm eff} \equiv \frac{8}{7} \left(\frac{11}{4}\right)^{4/3} \frac{\rho_{\rm dark}}{\rho_\gamma} \simeq 8.6 \left(1 - f_{\rm vis}\right) \alpha_S \left( \frac{g_{\star s}(T_{\rm dec})}{g_{\star s}(T_\star)} \right)^{1/3}.
\end{equation}
Here $g_{\star s}$ denotes the entropy degrees of freedom, with $g_{\star s}(T_{\rm dec})=43/11\simeq3.91$ after electron-positron annihilation, so that the numerical coefficient is $\tfrac{4}{7}(11/4)^{4/3}g_{\star s}(T_{\rm dec})\simeq8.6$. The ratio appears with the late-time value in the numerator because entropy release reheats the SM bath relative to the already-decoupled dark radiation, diluting an earlier injection.
 The observables $S_B$ and $\Delta N_{\rm eff}$ exhibit complementary scaling with the visible branching fraction $f_{\rm vis}$:
\begin{equation}
S_B \propto f_{\rm vis}, \qquad \Delta N_{\rm eff} \propto (1 - f_{\rm vis}).
\end{equation}
Visible-dominated reheating ($f_{\rm vis} \sim 1$) is severely constrained by primordial deuterium via $S_B$, whereas dark-dominated reheating ($f_{\rm vis} \sim 0$) is excluded by $\Delta N_{\rm eff}$~\cite{Planck:2018vyg,AtacamaCosmologyTelescope:2025nti}:
\begin{equation}
    \Delta N_{\rm eff}\lesssim\begin{cases}
        0.15\;(1\sigma),\;\, 0.30\;(2\sigma) & \text{Planck},\\
        0.07\;(1\sigma),\;\, 0.17\;(2\sigma) & \text{Planck}+\text{ACT}
    \end{cases}
\end{equation}
The $2\sigma$ entries are published one-tailed limits obtained by restricting to $\Delta N_{\rm eff}>0$, namely $\Delta N_{\rm eff}<0.30$ for Planck TT,TE,EE+lowE+lensing+BAO~\cite{Planck:2018vyg} and $\Delta N_{\rm eff}<0.17$ for P-ACT-LB~\cite{AtacamaCosmologyTelescope:2025nti}. The $1\sigma$ entries are not published and are estimated from the corresponding posteriors, $N_{\rm eff}=2.99\pm0.17$ and $N_{\rm eff}=2.86\pm0.13$, under the same restriction.
{Note that extra radiation cannot mimic $S_B$: the $\Delta N_{\rm eff}$ needed to raise $D/H$ would overshoot the measured $Y_p$~\cite{Poulin:2026ltf,Yeh:2022heq}.} The resulting restrictions on the DW and FOPT parameter spaces are shown in Fig.~\ref{fig:complementarity} for $D/H$ ratios computed with \texttt{PRIMAT} and \texttt{PArthENoPE}.

\begin{figure*}
    \centering
    \includegraphics[width=0.49\linewidth]{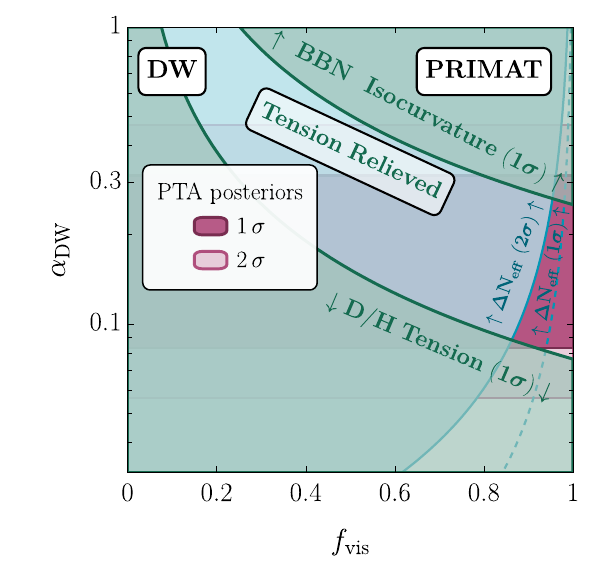}
    \includegraphics[width=0.49\linewidth]{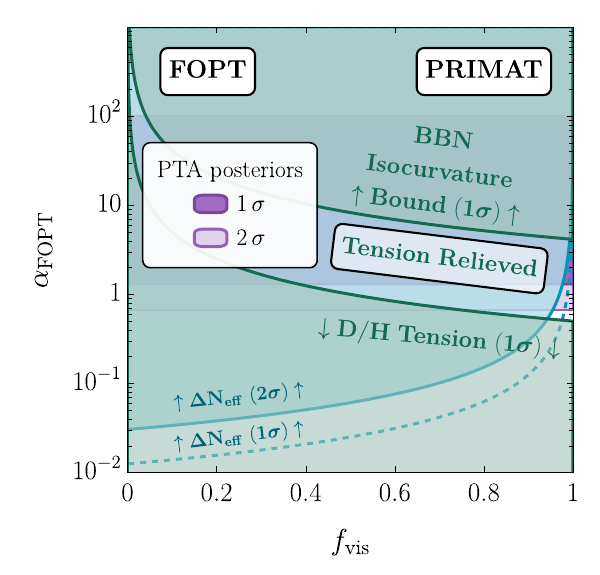}
    \includegraphics[width=0.49\linewidth]{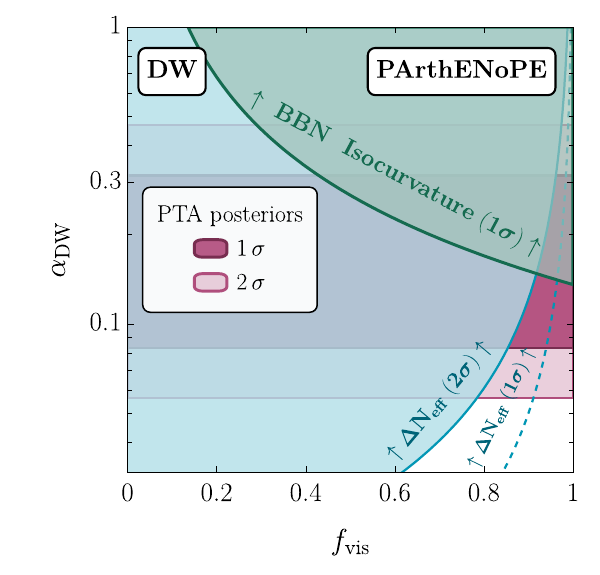}
    \includegraphics[width=0.49\linewidth]{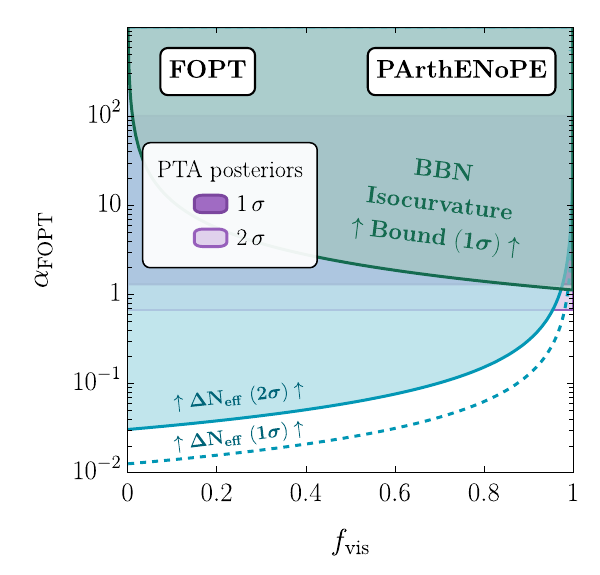}
   \caption{\itshape Complementarity of the BBN baryon-isocurvature and
    $\Delta N_{\rm eff}$ constraints as a function of the visible branching fraction
    $f_{\rm vis}$, for DWs (\textbf{left}, $\alpha_{\rm DW}$) and FOPTs (\textbf{right}, $\alpha_{\rm FOPT}$). \textbf{Top:} deuterium computed with \texttt{PRIMAT}, for which a finite range of $S_B$ alleviates the $D/H$ tension. \textbf{Bottom:} deuterium computed with \texttt{PArthENoPE}, for which the central values already agree, so only an upper bound on $S_B$ remains. In both cases the effect requires visible reheating, since $S_B\propto f_{\rm vis}$, and is therefore possible only for sufficiently large $f_{\rm vis}$. The $\Delta N_{\rm eff}$ bound scales as $\propto(1-f_{\rm vis})$ and excludes the dark-dominated region at small $f_{\rm vis}$. Shown are the Planck\,+\,ACT limits $\Delta N_{\rm eff}<0.17$ ($2\sigma$, solid) and $\Delta N_{\rm eff}<0.07$ ($1\sigma$, dashed) {for a reheating temperature of $T_{\rm reh}\simeq50 \rm \ MeV$}. The PTA $68\%$ ($1\sigma$) and $95\%$ ($2\sigma$) posteriors are independent of $f_{\rm vis}$ and therefore appear as horizontal bands.}
    \label{fig:complementarity}
\end{figure*}

\end{document}